\documentclass[journal]{IEEEtran}
\makeatletter
\def\ps@headings{%
\def\@oddhead{\mbox{}\scriptsize\rightmark \hfil \thepage}%
\def\@evenhead{\scriptsize\thepage \hfil \leftmark\mbox{}}%
\def\@oddfoot{}%
\def\@evenfoot{}}
\makeatother
\usepackage{amssymb,amsmath,amsfonts,bm,epsfig,graphicx,theorem}
\usepackage{rotating,setspace,latexsym,epsf,color,dsfont}
\usepackage{cite}

\newcommand{\Qv}{\mathbf{Q}}
\newcommand{\Uv}{\mathbf{U}}
\newcommand{\Vv}{\mathbf{V}}
\newcommand{\Pv}{\mathbf{P}}
\newcommand{\Tv}{\mathbf{T}}
\newcommand{\Sc}{\mathcal{S}}
\newcommand{\Dc}{\mathcal{D}}
\newcommand{\Nc}{\mathcal{N}}
\newcommand{\Tc}{\mathcal{T}}
\newcommand{\Ed}{\mathds{E}}

\newcommand{\lambdav}{\boldsymbol{\lambda}}
\newcommand{\epsilonv}{\boldsymbol{\epsilon}}

\newcommand{\tc}{\textcolor{black}}
\newtheorem{Theorem}{Theorem}
\newtheorem{Corollary}{Corollary}
\newtheorem{Lemma}{Lemma}

\newenvironment{Proof}[1]{\medskip\par\noindent
{\bf Proof:\,}\,#1}{{\mbox{\,$\blacksquare$}\par}}

\begin{document}

\IEEEoverridecommandlockouts

\title{Optimal Backpressure Scheduling in Wireless Networks using Mutual Information Accumulation\thanks{This work was supported in part by the Air Force Office of Scientific Research under grant FA9550-09-1-0140 and by the National Science Foundation under grant CCF 0963834.}}

\author{Jing~Yang,~\IEEEmembership{Member,~IEEE,}
        Yanpei~Liu,~\IEEEmembership{Student Member,~IEEE,}
        and~Stark~C.~Draper,~\IEEEmembership{Member,~IEEE}
\thanks{J. Yang, Y. Liu and S. C. Draper are with the Department
of Electrical and Computer Engineering, University of Wisconsin-Madison, WI, 53706, USA. e-mail: yangjing@ece.wisc.edu, yliu73@wisc.edu, sdraper@ece.wisc.edu.}}

\maketitle

\begin{abstract}
In this paper we develop scheduling policies that maximize the
stability region of a wireless network under the assumption that
mutual information accumulation is implemented at the physical
layer. When the link quality between nodes is not sufficiently high
that a packet can be decoded within a single slot, the system can
accumulate information across multiple slots, eventually decoding the
packet. The result is an expanded stability region. The accumulation
process over weak links is temporally coupled and therefore does not
satisfy the independent and identically distributed (i.i.d) assumption
that underlies many previous analysis in this area. Therefore the
problem setting also poses new analytic challenges. We propose two
dynamic scheduling algorithms to cope with the non-i.i.d nature of the
decoding. The first performs scheduling every $T$ slots, and approaches
the boundary of the stability region as $T$ gets large, but at the
cost of increased average delay. The second introduces virtual queues
for each link and constructs a virtual system wherein two virtual
nodes are introduced for each link. The constructed virtual system is
shown to have the same stability region as the original
system. Through controlling the virtual queues in the constructed
system, we avoid the non-i.i.d analysis difficulty and attain the full
stability region. We derive performance bounds for both algorithms and
compare them through simulation results.
\end{abstract}

\begin{IEEEkeywords}
Mutual information accumulation, backpressure algorithm, delay analysis, virtual queues.
\end{IEEEkeywords}

\section{Introduction}

Because of the broadcast nature of wireless communications, multiple
receiving nodes may overhear the transmission of a transmitting node
in a wireless network. Therefore, a packet can be routed through
different routes to its destination node. This ``multiple-route
diversity'' can be exploited to improve various measure of network
performance including throughput, average delay, and the probability
of transmission failure.

Different routing and scheduling policies have been proposed to
exploit the ``broadcast advantage'' of the wireless medium. Reference
\cite{ExOR} proposes ExOR, a packet-level opportunistic routing
protocol to exploit high loss-rate radio links in wireless
networks. Reference \cite{Rozner_2009} proposes a proactive link state
routing protocol that improves the forward nodes selection in
\cite{ExOR}. Reference \cite{Neely_2008} characterizes the maximum
network capacity region when packet-level opportunistic routing is
exploited, and proposes a routing algorithm named DIVBAR to stabilize
the system. DIVBAR adopts the idea of backpressure algorithms first
proposed in \cite{Tassiulas_1992, Tassiulas_1993}. References
\cite{ExOR, Rozner_2009, Neely_2008} discuss how to select a forwarder
based on the feedback information sent from the receivers that have
successfully received the packets. With a similar maximum weight
scheduling idea, \cite{Yeh_2007} analyzes the optimal information
theoretic based cooperative scheduling in a diamond network.  The
system is assumed to be able to adaptively select the optimal
encoding-decoding scheme such that any desirable transmission rate
vector within the corresponding capacity region can be achieved in
every slot. This ``fluid model'' is not quite practical. However,
because it is simple to analyze, it has been widely adopted in the
literature.

The policy space explored in the above papers, in particular those
having to do with wireless~\cite{ExOR, Rozner_2009, Neely_2008,
  Yeh_2007}, assumes that packet decoding is carried out within a
single slot.  This means that decoding depends only on current control
actions and channel quality.  However, there are many physical layer
techniques, such as energy accumulation and mutual information
accumulation (MIA), that do not fit this assumption.  Such techniques
do not require decoding to occur within a single slot but rather allow
the receivers to accumulation observations across multiple time slots
before decoding.  This allows the network to exploit weak radio links
more fully, thereby increasing system throughput.

In this paper we focus on mutual information accumulation and, while
there is some prior research on resource optimization in networks
using MIA \cite{Draper_2008,Urgaonkar_2010,mia_allerton11}, that work
does not consider questions of network stability and protocol design.
Our objectives in this work are therefore two fold.  First, we want to
characterize the maximum network capacity region when mutual
information accumulation is implemented at the physical layer. Second,
we aim to design joint routing and scheduling algorithms that
stabilize any point within the capacity region.






To make the concept of mutual information accumulation more tangible,
it helps to contrast it with energy accumulation.  In energy
accumulation multiple transmissions are combined non-coherently by
receiving nodes. This is usually enabled by using space-time or
repetition coding~\cite{Maric_2004, Maric_2005, Chen_2005}.  Mutual
information accumulation is more efficient, the difference between the
two being explained in~\cite{Draper_2008}. \tc{Consider a pair of senders transmitting information over two independent additive white Gaussian noise channels to the same receiver. Assume transmitter power $P$ and channel noise power $N$. Energy accumulation corresponds to the scenario when each transmitter sends the same codeword. When the decoder uses maximum ratio combining, a throughput of $\frac{1}{2}\log (1+\frac{2P}{N})$ bits per channel use can be achieved. With mutual information accumulation, independent parity symbols are sent, and the system can achieve $2\times\frac{1}{2}\log (1+\frac{P}{N})$ bits per channel use, which outperforms energy accumulation.} It has been noted in
\cite{Draper_2008} that for Gaussian channels at low signal-to-noise
ratio (SNR) energy accumulation is equivalent to mutual information
accumulation because capacity at low SNR is linear in SNR.  Mutual
information accumulation can be realized through the use of rateless
(or fountain) codes \cite{Draper_2008, Mitzenmacher_2004, LT_Code}.


Similar to \cite{ExOR, Rozner_2009, Neely_2008}, we assume our system
operates at packet-level, and each active transmitter transmits a
packet in each slot. Different from the probabilistic channel model
discussed in these references, we consider the scenario where the
channel state varies slowly, so that link conditions can be assumed to
be static over a long transmission period. Under this scenario, for a
weak link whose link rate is below the transmission rate, a packet
cannot be successfully delivered across the link in any single slot,
even with repetitive attempts in different slots. Thus, the
corresponding receiver can never decode and become a forwarder under
the schemes discussed in \cite{ExOR, Rozner_2009,
  Neely_2008}. However, when rateless codes are used by the
transmitters, although the corresponding receiver of the weak link
cannot decode the packet within a single slot, it can store that
corrupted packet and accumulate more information in later
slots. Eventually, a packet can be successfully delivered across a
weak link after a number of transmissions by accumulating information
across the slots. Thus, weak links can still be utilized in slowly
changing network environments. Compared with opportunistic routing
schemes, mutual information accumulation provides more reliable
throughput over weak links, and doesn't require the feedback
information to determine the forwarder in each slot. Thus it reduces
the required overhead.


Compared to the networks in \cite{Yeh_2007} and references \cite{cohen, Neely05}
where varying rate can be achieved within a slot through adaptive
encoding-decoding schemes, the mutual information accumulation scheme
doesn't require the encoding-decoding to be changed in every
slot. However, on average, it can achieve the same rate by repetitive
transmission and information bits accumulation. Therefore, the system
is more practical to implement without sacrificing throughput. On the
other hand, the information bit accumulation process also brings new
challenges to the design and analysis of routing and scheduling
algorithm.

The contribution of our work is three-fold.
\begin{itemize}
\item We characterize the maximum stability region of a wireless
  network enabled with mutual information accumulation under certain
  natural assumptions. Compared with networks where information
  accumulation is not allowed, the system is able to exploit weak
  links and an expanded stability region is thereby achieved.
\item We propose two dynamic routing and scheduling policies to
  achieve the maximum stability region. Both policies require simple
  coordination and limited overhead.
    \item The techniques we develop to cope with the temporally
      coupled mutual information accumulation process are novel and
      may have use in queuing problems more widely.
\end{itemize}

The rest of the paper is organized as follows. In
Section~\ref{sec:system}, we describe the system model. In
Section~\ref{sec:capacity}, we present the maximum stability region of
the network under our system model. In Section~\ref{sec:Tslot} and
Section~\ref{sec:virtual}, we design two different routing protocols
to achieve the maximum stability region and analyze the
performance. We present simulation evaluation in
Section~\ref{sec:simu}. Finally, we conclude in
Section~\ref{sec:conclusions}. Proofs are deferred to the appendices.

\section{System Model}\label{sec:system}
\subsection{The Basic System Setting}\label{sec:system_para}
We consider a time-slotted system with slots normalized to integral units	 $t\in\{0, 1, 2, \ldots\}$.	 There are $N$ network
nodes, and links are labeled according to {\it ordered} node pairs $(i,j)$ for $i,j \in \{1, \ldots , N\}$. We assume that there are $K$ different commodities in the network, $K\leq N$, where each commodity is labeled according to its destination node, e.g., all packets from commodity $c$ should be routed to destination node $c$. Data arrives randomly in packetized units. Let $A^c_i(t)$ denote the number of packets from commodity $c$ that exogenously arrive at network node $i$ during slot $t$.
Arrivals are assumed to be independent and identically distributed (i.i.d.) over timeslots, and we let $\lambda^c_i=E\{A^c_i(t)\}$ represent the arrival rate of commodity $c$ into source node $i$ in units of packets/slot. We assume $A^c_i(t)\leq A_{max}$ for all $c,i,t$.

We assume the channel fading states between any pair of nodes are stationary, and the active transmitters transmit with the same power level. Thus, a fixed reliable communication rate over an active link can be achieved in each slot. We expect that the algorithms developed in this paper can be modified to accommodate more general fading processes. 

\tc{We use $r_{ij}$ to denote the link rate between nodes $i,j$. We assume the link rate is reciprocal, i.e., $r_{ij}=r_{ji}$.}
We assume that each node can transmit at most one packet during any given node during any single time slot, i.e., $r_{ij}\leq 1$ packet/slot. We term the links with rate one packet per slot to be {\it strong} links; the rest of the links we term {\it weak} links. For weak links, we assume their rates are lower bounded by some constant value $r_{min}$, $0<r_{min}<1$. Define the set of neighbors of node $i$, $\Nc(i)$, as the set of nodes with $r_{ij}>0$, $j \in \Nc(i)$. We assume the size of $\Nc(i)$, denoted as $|\Nc(i)|$, is upper bounded by a positive integer $d$. We define $\mu_{max}$ as the maximum number of packets a node can successfully decode in any slot, which is equivalent to the maximum number of nodes that can transmit to a single node simultaneously. Therefore, we have $\mu_{max}\leq d$.

We assume the system operates under constraints designed to reduce interference among transmitters. Under the interference model, in any timeslot, only a subset of nodes are allowed to transmit simultaneously. We assume that transmissions from nodes active at the same time are interference free. We denote the set of feasible {\it activation pattern} as $\Sc$, where an activation pattern $s\in \Sc$ represents a set of active nodes. With a little abuse of the notation, we interchangeably use $s$ to represent an activation pattern and the set of active nodes of the pattern. \tc{For any $s\in \Sc$, we assume all of its subsets also belong to $s$. This allows us to assert that all of the nodes in an activation pattern always transmit when that activation pattern is selected.} This interference model can accommodate networks with orthogonal channels, restricted TDMA, etc.

\subsection{Mutual Information Accumulation at Physical Layer}\label{sec:system2}
 We assume mutual information accumulation is adopted at physical layer. Specifically, we assume that if a {\it weak} link with $r_{ij}<1$ is activated, rateless codes are deployed at its corresponding transmitter. When a packet transmitted over a weak link cannot be successfully decoded during one slot, instead of discarding the observations of the undecodable message, the receiver stores that partial information, and accumulates more information when the same packet is retransmitted. A packet can be successfully decoded when the accumulated mutual information exceeds the packet size. The assumption $r_{ij}>r_{min}$ implies that for any active link, it takes at most some finite number of time slots, $\lceil 1/r_{min}\rceil$, to decode a packet.

In order to simplify the analysis, we assume that $1/r_{ij}$ is an integer. The reason for this choice will become clear when our algorithm is proposed in Section~\ref{sec:capacity}. If $r_{ij}$ does not satisfy this assumption, we round it down to $1/\lceil 1/r_{ij}\rceil$ to enforce it.

The challenge in extending the backpressure routing framework to systems that use mutual information accumulation is illustrated through the following example. Suppose that a node $i$ has already accumulated half of the bits in packet 1 and half of the bits in packet 2. Since neither of the packets can be decoded, none of these bits can be transferred to the next hop, even though the total number of bits at node $i$ is equal to that of a full packet. This means that we need to handle the undecoded bits in a different manner. We also observe that, if node $i$ never accumulates enough information for packets 1 or 2, then these packets can never be decoded and will be stuck at node $i$. If we assume that the system can smartly drop the undecoded partial packet whenever a fully decoded copy is successfully delivered to its destination, coordination among the nodes is required to perform this action. The overhead of the coordination might offset the benefits brought by mutual information accumulation. Moreover, unlike the opportunistic model studied in \cite{ExOR, Rozner_2009, Neely_2008}, given that weak link is active, whether or not a successful transmission will occur in that slot is not an i.i.d random variable but rather a deterministic function of the number of already accumulated bits of that packet at the receiving node. This difference makes the analysis even more complicated.


Therefore, in the following sections, we define two different types of queues. One is the traditional {\em full packet} queue which stores fully decoded packets at each node. The other type of queue is a {\em partial packet} queue. It represents the fraction of accumulated information bits from a particular packet. The specific definition of the queues and their evolution depends on the scheduling policy, and will be clarified in Section~\ref{sec:Tslot}, and Section~\ref{sec:virtual}, respectively.

We assume each node has infinite buffer space to store fully decoded packets and partial packets. Overflow therefore does not occur.

\subsection{Reduced Policy Space}
The policy space of the network we consider can be much larger than that of networks without mutual information accumulation. First, because of the broadcast nature of wireless communication, multiple receivers can accumulate information regarding the same packet in any given slot, and a receiver can collect information on a single packet from multiple nodes. Keeping track of the locations of different copies of the same packet requires a lot of overhead. Second, allowing multiple receivers to store copies of the same packet introduces more traffic into the network. Therefore, stabilizing the network requires a sophisticated centralized control strategy. Finally, accumulating information bits of a packet from multiple nodes makes the decoding options of a packet increase exponentially; thus characterizing the network maximum stability region becomes intractable. Therefore, we make the following assumptions:
\begin{itemize}
{\it \item[A1.] For each packet in the system, at any given time, only one node is allowed to keep a fully decoded copy of it.
\item[A2.] In addition to the node with the fully decoded copy, one other node is chosen as the potential forwarder for the packet. Only the potential forwarder is allowed to accumulate information about the packet.}
\end{itemize}

Restricting ourselves to this policy space may sacrifice some part of the stability region that could be achieved under a more general policy space. However, as we will see, these assumptions enable us to explicitly characterize the maximum stability region with the given policy space. Compared with systems that operate without the capability to accumulate mutual information, our system is able to exploit weak links even when one-slot decoding is not possible. The stability region will thereby be greatly enlarged.

If a node directly contributes to the successful decoding of a packet at another node, this node is denoted as a {\it parent} for that packet. Assumptions A1-A2 guarantee that for any packet in the network, there is only one parent at any given time.
We also note that, if we relax assumptions A1-A2, and make the following assumption instead
\begin{itemize}
{\it \item[A3.] Every packet existing in the network has a single parent at any given time. In other words, the accumulated information required to decode a packet at a node is received from a single transmitting node.}
\end{itemize}
then, the maximum stability region under A1-A2 and that under A3 are equivalent. Assumption A3 means that we don't allow a node to accumulate information from different nodes (different copies of the same packet) to decode that packet. However, multiple copies of a packet may still exist in the network.

\section{Network Capacity with Mutual Information Accumulation}\label{sec:capacity}
In this section we characterize the optimal throughput region under all possible routing and scheduling algorithms that conform to the network structure specified in Section~\ref{sec:system_para} and Assumption A3.

At the beginning of each slot, a certain subset $s$ of nodes is selected to transmit in the slot. Any node $i\in s$ can transmit any packet that it has received {\it and} successfully decoded in the past. Note that because packets must be decoded prior to transmission, partial packets cannot be delivered. For node $j\in\Nc(i)$, if it has already decoded a packet being transmitted, it can simply ignore the transmission; otherwise, it listens to the transmission and aims to decode it at the end of that slot. Receiving nodes connected with strong links can decode the packet in one slot. Nodes connected with weak links cannot decode the packet in one slot. Rather, they need to listen to the same node transmitting the same packet over a number of slots.
A packet is said to be successfully delivered to its destination node when the {\it first copy} of the packet is decoded by its destination node. These assumptions allow for any possible routing and scheduling policy satisfying Assumption A3. \tc{We note that packets arrive at a node from a single commodity may get delivered to their destination node in a permuted order since each packet may take a different route.}

Let $\lambdav$ represent the input rate vector to the system, where $\lambda^c_i$ is the input rate of commodity $c$ entering node $i$. Define $Y^c_i(t)$ as the number of packets from commodity $c$ that originated at node $i$ and have been successfully delivered to destination node $c$ over $[0,t)$. According to the definition of network stability \cite{Neely_now}, a policy is defined as {\it stable} if
    \begin{align}
    \lim_{t\rightarrow \infty} \frac{Y^c_i(t)}{t}=\lambda^c_i,\qquad \forall c.
    \end{align}

Stronger definitions of stability can be found in \cite{Tassiulas_1992, Tassiulas_1993, stable94}.
The maximum stability region or {\it network layer capacity region} $\Lambda$ of a wireless network with mutual information accumulation is defined as the closure of all $\lambdav$ that can be stabilized by the network according to some policy with the structure described above.

\begin{Theorem}\label{thm1}
For a network with given link rates $\{r_{ij}\}$ and a feasible activation pattern set $\Sc$,
the network capacity region $\Lambda$ under assumption A3 consists of all rate vectors $\{\lambda^c_n\}$ for which there exists flow variables $\{\mu^c_{ij}\}$ together with a probability $\pi_s$ for each possible activation pattern $s\in \Sc$ such that
\begin{align}
\mu^c_{ij}&\geq 0,\quad \mu^c_{ci}=0,\quad \mu^c_{ii}=0, \quad \forall i,j,c\label{eqn:cap1}\\
\sum_{l}\mu^c_{li}+\lambda^c_i&\leq \sum_{j}\mu^c_{ij},\quad \forall i\neq c, \forall c\label{eqn:cap2}\\
\sum_c\mu^c_{ij}&\leq \sum_c\sum_{s\in \Sc}\pi_s\theta^c_{ij}(s)r_{ij},\quad \forall i,j\label{eqn:cap3}\\
\sum_{s\in \Sc}\pi_s&\leq1,
\end{align}
where the probabilities $\theta_{ij}(s)$ satisfy
\begin{align}
\theta^c_{ij}(s)&=0 \mbox{ \rm{if} }i\notin s, \\
\quad \sum_c\sum_{j}\theta^c_{ij}(s)&= 1,\forall i.\label{eqn:cap5}
\end{align}
\end{Theorem}

The necessity of this theorem can be proved following the same approach in \cite{Neely_2008} and is provided in Appendix~\ref{apx:thm1}. The sufficiency part is proved in Section~\ref{sec:Tslot} by constructing a stabilizing policy for any rate vector $\lambdav$ that is in the interior of capacity region.

The capacity region is essentially similar to the capacity theorem of \cite{Neely_now,Neely_2008}. The relations in (\ref{eqn:cap1}) represent non-negativity and flow efficiency constraints for conservation constraints. Those in (\ref{eqn:cap2}) represent flow conservation constraints. Those in (\ref{eqn:cap3}) represent link constraints for each link $(i,j)$. The variable $\theta^c_{ij}(s)$ can be interpreted as the probability that the transmission over link $(i,j)$ eventually contributes to the delivery of a packet of commodity $c$ at node $c$, given that the system operates in pattern $s$. In other words, link $(i,j)$ is on the routing path for this packet from its origin node to its destination node.

This theorem implies that the network stability region under Assumption A3 can be defined in terms of an optimization over the class of all stationary policies that use only single-copy routing. Thus, for any rate vector $\lambdav\in \Lambda$, there exists a stationary algorithm that can support that input rate vector by single-copy routing all data to the destination.

The $\Lambda$ defined above are in the same form as when a ``fluid model'' is considered. In other words, the extra decoding constraint imposed by mutual information accumulation does not sacrifice any part of the stability region. We can simply ignore the packetization effect when we search for the maximum stability region.

The $\sum_c\mu^c_{ij}$ defining the stability region represents the {\it effective} flow rate over link $(i,j)$. An ``effective'' transmission means that the bits transferred by that transmission eventually get delivered to the destination. If the transferred information becomes a redundant copy, or discarded partial packet, the transmission is not effective, and doesn't contribute to the effective flow rate. We can always make the inequalities (\ref{eqn:cap2})-(\ref{eqn:cap3}) tight by controlling the value of $\pi_s$.

Solving for the parameters $\{\pi_s\}$ and $\{\theta^c_{ij}(s)\}$ required to satisfy the constraints requires a complete knowledge about the set of arrival rates $\{\lambda^c_i\}$, which cannot be accurately measured or estimated in real networks. On the other hand, even when $\lambdav$ is given, solving the equations can still be quite difficult. In the following, we overcome this difficulty with online algorithms which stabilize any $\lambdav$ within $\Lambda$, but with a possibly increased average delay as $\lambda$ approaches the boundary of $\Lambda$.

\section{$T$-slot Dynamic Control Algorithm}\label{sec:Tslot}
In the following, we construct a policy that fits Assumptions A1-A2. Although more restrictive than Assumption A3, we will see that the stronger assumptions do not compromise stability performance in the sense that they do not reduce the stability region.
To construct a dynamic policy that stabilizes the system anywhere in the interior of $\Lambda$, specified in Theorem~\ref{thm1}, we first define our decision and queues variables.

We assume that each packet entering the system is labeled with a unique index $k$. At time $t$, $0\leq k\leq \sum_{c,i}\sum_{\tau=1}^t A^c_i(\tau)$. If packet $k$ belongs to commodity $c$, we denote it as $k\in \Tc_c$. Let $\left\{\beta_{ij}^{(k)}(t)\right\}$  represent the binary control action of the system at time $t$. Specifically, $\beta_{ij}^{(k)}(t)=1$ means that at time $t$, node $i$ transmits packet $k$ to node $j$. We restrict the possible actions so that in each slot each node transmits at most one packet, i.e.,
\begin{align}\label{beta_con}
 &\sum_{j,k}\beta_{ij}^{(k)}(t)\leq 1,\quad \forall i,
\end{align}
and at most one node is chosen as the forwarder for packet $k$, i.e.,
\begin{align}\label{beta_con2}
 \sum_{i,j}\beta_{ij}^{(k)}(t)\leq 1, \quad\forall k.
\end{align}

Because of the mutual information accumulation property, even if packet $k$ is transmitted over link $(i,j)$ in slot $t$, it doesn't necessarily mean that packet $k$ can be decoded at node $j$ at the end of slot $t$. In particular, under the fixed link rate assumption, successful transmission cannot occur over weak links in a single timeslot.
We let $f_{ij}^{(k)}(t)$ be an indicator function where $f_{ij}^{(k)}(t)=1$ indicates that the packet $k$ has been successfully delivered from node $i$ to node $j$ in slot $t$. The indicator function is a function of the current control action and partial queue status at the beginning of slot $t$. Apparently, $f_{ij}^{(k)}(t)=1$ implies that $\beta_{ij}^{(k)}(t)=1$.

As discussed in Section~\ref{sec:system2}, we define two types of queues at each nodes.
One is to store the fully received and successfully decoded packet at the nodes, while the other queue stores the partially received packets.
We use $Q^c_i(t)$ to denote the length of node $i$'s queue of fully received packets from commodity $c$ at time $t$, and use $P_i^{(k)}(t)$ to represent the total fraction of packet $k$ accumulated by node $i$ up to time $t$. The sum-length of partial queues of commodity $c$ at node $i$ storing partial packets can be represented as $P^c_i(t)=\sum_{k\in \Tc_c}P_i^{(k)}(t)$. The fraction of packet $k$, $P_i^{(k)}(t)$, can be cleared either when packet $k$ is successfully decoded and enters the full packet queue $Q^c_i$, or when the system controller asks node $i$ to drop packet $k$.  \tc{With a little abuse of notation, we use $Q_i^c$ and $P_i^c$ to denote the full packet queue and partial packet queue from commodity $c$ at node $i$, respectively.}

Then, according to our Assumptions A1-A2, the queue lengths evolve according to
\begin{align}
Q^c_i(t+1)&=\Big( Q^c_i(t)-\sum_{j,k\in \Tc_c}\beta_{ij}^{(k)}(t)f^{(k)}_{ij}(t)\Big)^+\nonumber\\
&\quad+\sum_{l,k\in \Tc_c}\beta_{li}^{(k)}(t)f^{(k)}_{li}(t)+A^c_i(t)\\
P_i^{(k)}(t+1)&=P_i^{(k)}(t)+\sum_{l}\beta_{li}^{(k)}(t)r^{(k)}_{li}(t)\hspace{-0.02in}-\hspace{-0.03in}\sum_{l}\beta_{li}^{(k)}(t)f^{(k)}_{li}(t)\nonumber\\
&\quad-\sum_{l,(m\neq i)}P_i^{(k)}(t)\beta_{lm}^{(k)}(t)\label{eqn:p1}
\end{align}
where
\begin{align}
r^{(k)}_{li}(t)&=\left\{\begin{array}{ll}r_{li}&  P_i^{(k)}(t)+r_{li}\leq 1\\
1-P_i^{(k)}(t)& P_i^{(k)}(t)+r_{li}> 1
\end{array}\right.
\end{align}
and $(x)^+=\max\{x,0\}$.
Under the assumption that $1/r_{ij}$ is an integer for every $(i,j)$, we have $r^{(k)}_{li}(t)=r_{li}$.

Since we only allow there to be only a forwarder for any given packet at any time, if $\beta_{lm}^{(k)}(t)=1$, any nodes other than node $m$ which have accumulated partial information of packet $k$ must drop that partial packet $k$. This effect results in the last negative term in (\ref{eqn:p1}). On the other hand, the first negative term in (\ref{eqn:p1}) corresponds to successful decoding of packet $k$, after which it is removed and enters $Q^c_i$ for some $c$.

\subsection{The $T$-slot Algorithm}
Our algorithm works across epochs, each consisting of $T$ consecutive timeslots. Action decisions are made at the start of each epoch and hold constant through the epoch. We analyze the choice of $T$ on the stability region and average backlog. Any rate vector $\lambdav$ inside the capacity region $\Lambda$ can be stabilized by a sufficiently large choice of $T$.

\begin{itemize}
\item[1)] \textbf{Check single-link backpressure.} At the beginning of each epoch, i.e., when $t=0,T,2T,\ldots$, node $i$ checks its neighbors and computes the differential backlog weights
$$W_{ij}(t)=\max_c[Q^c_i(t)-Q^c_j(t)]^+r_{ij},\quad j\in\Nc(i) .$$
Denote the maximizing commodity as $$c^*_{ij}=\arg \max_c [Q^c_i(t)-Q^c_j(t)]^+.$$
\item[2)] \textbf{Select forwarder.} Choose the potential forwarder for the packets in $Q_i$ as the node $j$ with the maximum weight $W_{ij}(t)$. Denote this node as $j^*_i=\arg \max_j W_{ij}(t)$.
\item[3)] \textbf{Choose activation pattern.} Define the activation pattern $s^*$ as the pattern $s\in S$ that maximizes
$$\sum_{i\in s}W_{ij^*_i}.$$
Any node $i\in s^*$ with $W_{ij^*_i}>0$ transmits packets of commodity $c^*_{ij^*_i}$ to $j^*_i$. The pairing of transmitter $i\in s^*$ and receiver $j^*_i$ and the commodity being transmitted $c^*_{ij^*_i}$ is continued for $T$ consecutive timeslots.
\item [4)] \textbf{Clear partial queues.} Release all the accumulated bits in the partial queue $P^c_i$, $\forall i,c$, at the end of each epoch.
\end{itemize}
The $T$-slot algorithm satisfies constraints (\ref{beta_con})-(\ref{beta_con2}). The ``potential forwarder'' in Step 2) refers to the forwarder of node $i$ if node $i$ is active.
We clear all of the partial queues in the system every $T$ slots (in Step 4)) for the simplicity of analysis. It is likely not the best approach to handle the partial queues. Intuitively, the performance should be improved if we only release the partial queues when a selected forwarder for a packet is not the same as the previous one (thus satisfying A3).


\begin{Theorem}\label{thm:Tslot}
The algorithm stabilizes any rate vector satisfying $\lambdav+\epsilonv(T) \in \Lambda$, where $\epsilonv(T)$ is a vector with minimum entry $\epsilon> 1/T$. The average expected queue backlog $\lim_{t\rightarrow \infty}\frac{1}{t}\sum_{\tau=0}^{t-1}\sum_{c,i}\Ed\{Q^c_i(\tau)\}$ in the system is upper bounded by $$\frac{KNT^2(\mu_{max}\hspace{-0.02in}+\hspace{-0.02in}A_{max})^2+NT^2}{2(\epsilon T-1)}+\frac{KN(T\hspace{-0.02in}-\hspace{-0.02in}1)(\mu_{max}\hspace{-0.02in}+\hspace{-0.02in}A_{max})}{2}.$$
\end{Theorem}
The proof of Theorem~\ref{thm:Tslot} is provided in Appendix~\ref{apx:thm_Tslot}. The proof is based on the fact that the $T$-slot algorithm minimizes the $T$-slot Lyapunov drift, which is shown to be negative when $\sum_{c,i}Q^c_i(t)$ is sufficiently large.

The constructed algorithm proves the sufficiency of Theorem~\ref{thm1}. The intuition behind the algorithm is that by using the weak links consecutively over a long window, the potentially contributing loss caused by dropped partial packets is kept small, therefore, the effective rates over the weak links can get close to the link rate. The algorithm approaches the boundary of the capacity region in $O(1/T)$.

When $T$ is large enough, the average expected backlog in the system scales as $O(T)$. For this reason, in the next section, we introduce a virtual queue based algorithm which updates action every single slot. We expect that average backlog under the virtual queue based algorithm will be improved since its upper bound does not scale as $O(T)$.

Given $T> \frac{1}{\epsilon}$, the upper bound \tc{in Theorem~\ref{thm:Tslot}} is a convex function of $T$.
This implies that for any $\lambdav+\epsilonv(T)\in \Lambda$, there exists an optimal value of $T$ which stabilizes the system and introduces minimal delay bound. However, when the arrival rates are unknown, it may not be practical to search for this optimal value.

Finally, we note that for some special values of $T$, the network can still be stabilized even when $T\leq 1/\epsilon$. For example, when $T$ is chosen as $\prod_{(i,j)}\frac{1}{r_{ij}}$, then, under any possible activation pattern $s\in\Sc$, all partial packets are decoded at the end of the $T$-slot window. This implies that the policy can stabilize any $\lambdav+\epsilonv\in\Lambda$. This phenomena will be illustrated through examples in Section~\ref{sec:simu}. For small networks, such a value for $T$ can be easily computed and may be small; for large networks with many weak links, such value may still be quite large.

\section{Virtual Queue Based Algorithm}\label{sec:virtual}

In this section, we develop a second algorithm that exhausts the stability region without needing to set a large $T$. Thereby it attains better delay performance. As seen in Theorem~\ref{thm1}, the delay is caused by the long time window of planning and infrequent update of the control action. Therefore, in order to obtain better delay performance, intuitively, we need to update our policy more frequently. This requires us to design more sophisticated mechanisms to handle the partial packet queues and additional analysis challenges brought by the temporally coupled decoding process over weak links.

Our approach is to construct a network that contains ``virtual'' queues, which handle the partial packets and decoding process over weak links. The resulting network has the same maximum stability region as the original network. By stabilizing the constructed network, the original network is also stabilized.

Specifically, in order to handle the partial packet queue in a simple and effective way, we introduce buffers over weak links. We assume there is a buffer at the transmitter side for each weak link. Then, if a node wants to send a packet over a weak link, the packet is pushed into the buffer. The buffer keeps the packet until it is successfully decoded at the corresponding receiver.
The intuition behind the introduction of these buffers is that, since we don't want dropped partial packets to lose much effective rate over weak links, once the system picks a forwarding node for a particular packet, the system never changes this decision.

For transmissions over weak links, a packet can only be decoded and transferred to next hop when enough information is collected. Under the $T$-slot algorithm, since control actions updates every $T$ slot, and partial queues are cleared at the end of every epoch, it is relatively simple to track the queue evolution and perform the analysis. When control actions change every slot, queues may evolve in a more complicated way and thus difficult to track and analyze.
In order to overcome the analysis challenges, we introduce a second buffer at the receiver side of each weak link. Under the proposed algorithm, we ensure that the receiver never tries to decode a packet until it accumulates enough information, i.e., queue length in the second buffer only decreases when it is sufficiently long. By doing this, the evolution of the queue lengths can be tracked and analyzed easily.

\tc{Essentially, we only need to introduce virtual nodes and virtual buffers over weak links in order to handle partial packets. However, link rates may vary over time and vary in different activation patterns (discussed in Sec.~\ref{sec:vary}). Therefore, for the virtual queue based algorithm, we introduce virtual nodes and buffers over both weak links and strong links, and treat them uniformly.}

\subsection{The Virtual Queue Vector}
We divide $Q^c_i(t)$ into two parts. The first stores the packets that have not yet been transmitted in any previous slots, denoted as $U^c_i(t)$. The second stores packets partially transmitted over some links but not yet decoded, denoted as $V^c_i(t)$. Since each packet in the second part is associated with some link, in order to prevent any loss of effective rate caused by dropped partial packets, we require these packets to be transmitted over the same link until they are decoded. 
 We use $V^{(k)}_{ij}(t)$ to denote the information of packet $k$ required to be transmitted over link $(i,j)$, and $P^{(k)}_{ij}(t)$ to denote the accumulated information of packet $k$ at node $j$.
We define $V^c_{ij}(t)=\sum_{k\in \Tc_c} V^{(k)}_{ij}(t)$, and $P^c_{ij}(t)=\sum_{k\in \Tc_c} P^{(k)}_{ij}(t)$, where we recall that $\Tc_c$ is the set of packets of commodity $c$. Note that $P^c_{ij}(t)$ is different from $P^c_{j}(t)$ defined in Section~\ref{sec:Tslot}, since the latter is associated with node $j$ and the former is associated with link $(i,j)$.

\begin{figure}[t]
\begin{center}
\scalebox{0.45} {\epsffile{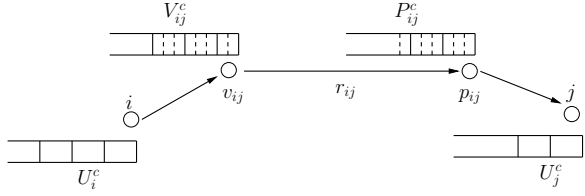}}
\end{center}
\vspace{-0.15in}
\caption{The constructed virtual system.}
\label{fig:virtual}
\vspace{-0.15in}
\end{figure}

Associated with virtual queues, we define {\it virtual nodes}, as depicted in Fig.~\ref{fig:virtual}. For the link from node $i$ to node $j$, we associate one virtual node with $\{V^c_{ij}\}_c$ and a second with $\{P^c_{ij}\}_c$. The virtual node associated with $\{V^c_{ij}\}_c$ is denoted as $v_{ij}$, while the virtual node associated with $\{P^c_{ij}\}_c$ is denoted as $p_{ij}$. We have decomposed the weak link $(i,j)$ into three links: $(i,v_{ij}), (v_{ij},p_{ij}),(p_{ij},j)$, with link rates $1,r_{ij},1$, respectively. The virtual nodes and corresponding link rates for link $(j,i)$ can be defined in a symmetric way.

We follow the definition of control actions, where $\left\{\beta_{ij}^{(k)}(t)\right\}$ represent the control action of the system at time $t$. Depending on whether or not packet $k$ has already been transmitted by node $i$, we also divide the decision actions into two types, denoted as $\beta^{1(k)}_{ij}$ and $\beta^{2(k)}_{ij}$, respectively.
In the following algorithm, we only make control decisions for the packets at the head of corresponding queues. Therefore we can replace the superscript packet index $(k)$ by its commodity $c$ without any worry of confusion.

When $\beta^{1c}_{ij}(t)=1$, node $i$ pushes a {\it new} packet from $U^c_i(t)$ into the tail of $V^c_{ij}$ at the beginning of slot $t$. This implies that the system assigns node $j$ to be the next forwarder for that packet. Once the packet is pushed into $V^c_{ij}$, we transmit the packet that is at the head of $V^c_{ij}$ to node $j$, generally a different packet. Thus, an amount $r_{ij}$ of information can be accumulated at the tail of $P^c_{ij}(t)$, and the length of $V^c_{ij}$ is reduced by $r_{ij}$. This mechanism ensures that the packets in the virtual buffer is transmitted and decoded in a FIFO fashion.

When $\beta^{2c}_{ij}(t)=1$, without pushing a new packet into the buffer, we retransmit the packet at the head of $V^c_{ij}(t)$.
We let
\begin{align}
\beta^c_{ij}(t)&=\beta_{ij}^{1c}(t)+\beta^{2c}_{ij}(t).
\end{align}
We require that
\begin{align}\label{eqn:bcon}
\sum_{c,j}\beta^c_{ij}(t)\leq 1,\quad \forall i,t.
\end{align}

Further, we define $f^c_{ij}(t)\in\{0,1\}$ as binary decoding control actions. $f^c_{ij}(t)=1$ indicates that receiver $j$ has accumulated enough information to decode the packet at the head of $P^c_{ij}(t)$. It then moves that packet out of $P^c_{ij}(t)$ and into $U^c_j(t)$. We impose the following constraint
\begin{align}\label{eqn:fcon}
f^c_{ij}(t)&\leq P^c_{ij}(t),\quad\forall c,i,j
\end{align}
which indicates that $f^c_{ij}(t)=1$ only when $P^c_{ij}(t)\geq 1$, i.e., receiver $j$ has accumulated enough information to decode the packet at the head of $P^c_{ij}(t)$. The reason for imposing this constraint on $f^c_{ij}(t)$ is that, we cannot simply use $(P^c_{ij}(t)-f^c_{ij}(t))^+$ to represent the queue length of $P^c_{ij}$ after a decoding action is taken at that queue. If $P^c_{ij}(t)<1$, even if $f^c_{ij}(t)=1$, after the decoding action, the queue length will still be $P^c_{ij}(t)$ since no packet can be successfully decoded. This is not equal to $(P^c_{ij}(t)-f^c_{ij}(t))^+$, which is zero in this scenario.

Then, according to constraints (\ref{eqn:bcon}) and (\ref{eqn:fcon}), the queue lengths evolve according to
\begin{align}
U^c_i(t+1)&=\Big( U^c_i(t)\hspace{-0.03in}-\hspace{-0.03in}\sum_{j}\beta_{ij}^{1c}(t)\Big)^+\hspace{-0.03in}+\hspace{-0.03in}\sum_{l}f^c_{li}(t)\hspace{-0.03in}+\hspace{-0.03in}A^c_i(t)\label{eqn:u}\\
V^c_{ij}(t+1)&\leq\left(V^c_{ij}(t)+\beta_{ij}^{1c}(t)(1-r_{ij})-\beta^{2c}_{ij}(t)r_{ij}\right)^+\nonumber\\
&= \left(V^c_{ij}(t)+\beta_{ij}^{1c}(t)-\beta^c_{ij}(t)r_{ij}\right)^+\label{eqn:v}\\
P^c_{ij}(t+1)&\leq P^c_{ij}(t)+\beta^c_{ij}(t)r_{ij}-f^c_{ij}(t)\label{eqn:p}
\end{align}
The inequalities in (\ref{eqn:v}) and (\ref{eqn:p}) come from the fact that $\beta^{1c}_{ij}(t)$ and $\beta^{2c}_{ij}(t)$ can be applied to a dummy packet when a queue is empty. When the corresponding queue is not empty, the inequality becomes an equality.

Define
\begin{align}\label{eqn:lya12}
L(\Uv(t),\Vv(t),\Pv(t))&=\sum_{c,i}(U^c_i(t))^2+\sum_{c,(i,j)}\hspace{-0.03in}(V^c_{ij}(t))^2\nonumber\\
&\quad+\sum_{c,(i,j)}\hspace{-0.03in}(P^c_{ij}(t)-\eta)^2
\end{align}
where $\eta$ is a parameter used to control the length of $P^c_{ij}(t)$.
Define $\Delta(t)$ as the one-slot sample path Lyapunov drift:
\begin{align*}
\Delta(t)&:= L(\Uv(t+1),\hspace{-0.03in}\Vv(t+1),\hspace{-0.02in}\Pv(t+1))-L(\Uv(t),\hspace{-0.03in}\Vv(t),\hspace{-0.02in}\Pv(t))
\end{align*}

\begin{Lemma}\label{lemma:1drift}
Under constraints (\ref{eqn:bcon}) and (\ref{eqn:fcon}), the sample path Lyapunov drift satisfies
\begin{align}
\Delta(t)&\leq 2\sum_{c,i}U^c_i(t)A^c_i(t)-2\sum_{c,i,j}[U^c_i(t)-V^c_{ij}(t)]\beta^{1c}_{ij}(t)\nonumber\\
&\quad-2\sum_{c,i,j}[V^c_{ij}(t)-P^c_{ij}(t)]r_{ij}\beta^c_{ij}(t)\nonumber\\
&\quad -2\sum_{c,i,j}[P^c_{ij}(t)-\eta-U^c_j(t)]f^c_{ij}(t)+\alpha_2\label{eqn:delta}
\end{align}
where
\begin{align}
\alpha_2&=KN(d+A_{max})^2+2N+KNd
\end{align}
\end{Lemma}
The proof of this lemma is provided in Appendix~\ref{apx:lemma_1drift}.

\subsection{The Algorithm}\label{sec:virtual_algo}
In contrast to the algorithm of Section~\ref{sec:Tslot}, this algorithm updates every timeslot. The purpose of the algorithm is to minimize the right hand side of (\ref{eqn:delta}) given the current $\Uv,\Vv,\Pv$.

\begin{itemize}
\item[1)] \textbf{Find per-link backpressure.} At the beginning of a timeslot, node $i$ checks its neighbors and computes the differential backlogs. We compute the weight for the link between node $i$ and the first virtual node, and, separately, the weight for the link between the two virtual nodes. Specially, the weight for control action $\beta^{1c}_{ij}$ is computed as
$$W^{1c}_{ij}(t)= [U^c_i(t)-V^c_{ij}(t)+(V^c_{ij}(t)-P^c_{ij}(t))r_{ij}]^+$$
and the weight for control action $\beta^{2c}_{ij}$ is computed as
$$W^{2c}_{ij}(t)=[V^c_{ij}(t)-P^c_{ij}(t)]^+r_{ij}$$
The weight of commodity $c$ over link $(i,j)$ is $W^c_{ij}(t)=\max\{W^{1c}_{ij}(t),W^{2c}_{ij}(t)\}$.
The weight for the link $(i,j)$ is $$W_{ij}(t)=\max_c W^{c}_{ij}(t),$$ and the optimal commodity $$c^*_{ij}=\arg\max_c W^{c}_{ij}(t).$$
\item[2)] \textbf{Select forwarder.} Choose the potential forwarder of the current slot for node $i$ with the maximum weight $W_{ij}(t)$ and denote it as  $j^*_i=\arg \max_j W_{ij}(t)$.
\item[3)] \textbf{Choose activation pattern.} Define the optimal activation pattern $s^*$ as the pattern $s\in S$ that maximizes
$$\sum_{i\in s}W_{ij^*_i}.$$
\item[4)] \textbf{Transmit packets.} For each $i\in s^*$, if $W_{ij^*_i}>0$, let node $i$ transmit a packet of commodity $c^*_{ij^*_i}$ to node $j^*_i$. For strong links, node $i$ transmits a packet from the head of $U^{c^*}_i$. If link $(i,j^*)$ is a weak link, and $W_{ij}(t)=W_{ij}^1(t)$, node $i$ pushes a new packet from $U^{c^*}_i$ into $V^{c^*}_{ij^*_i}$ and transmits the packet from the head of $V^{c^*}_{ij^*_i}(t)$; otherwise, node $i$ resends the packet at the head of $V^{c^*}_{ij^*_i}(t)$.
\item[5)] \textbf{Decide on decoding actions.} For each link $(i,j)$ and each commodity $c$, we choose $f^c_{ij}(t)\in\{0,1\}$ to maximize
\begin{align}
[P^c_{ij}(t)-\eta-U^c_{j}(t)]f^c_{ij}(t)
\end{align}\tc{where $\eta$ is a parameter greater than or equal to 1. We let $f^c_{ij}(t)=1$ when $P^c_{ij}(t)-\eta-U^c_{j}(t)=0$.}
\end{itemize}

\begin{Lemma}\label{lemma:Pij}
Under the above virtual queue based algorithm: $(a)$ If $P^c_{ij}(t)<\eta$ for some weak link $(i,j)$ and slot $t$, then $f^c_{ij}(t)=0$. $(b)$ If $P^c_{ij}(t_0)\geq \eta-1$, then, under the proposed algorithm, $P^c_{ij}(t)\geq \eta-1$ for every $t\geq t_0$.
\end{Lemma}
\begin{Proof}
In order to maximize $(P^c_{ij}(t)-\eta-U^c_j(t))f^c_{ij}(t)$, $f^c_{ij}(t)=1$ only when $P^c_{ij}(t)-\eta-U^c_j(t)>0$. Therefore, if $P^c_{ij}(t)<\eta$, $f^c_{ij}(t)$ must equal zero, which proves $(a)$.

Now suppose that $P^c_{ij}(t)\geq \eta-1$ for some slot $t$. We show that it also holds for $t+1$. If $P^c_{ij}(t)\geq \eta$, then it can decrease by at most one packet on a single slot, so that $P^c_{ij}(t+1)\geq P^c_{ij}(t)-f^c_{ij}(t)\geq \eta-1$. If $P^c_{ij}(t)<\eta$, we must have $f^c_{ij}(t)=0$, the queue cannot decrease in slot $t$, and we again have $P^c_{ij}(t+1)\geq \eta-1$.
\end{Proof}

With Lemma~\ref{lemma:Pij}, we can see that when setting $\eta=1$, under the proposed algorithm, if $P^c_{ij}(t)<1$ for some weak link $(i,j)$ and slot $t$, then $f^c_{ij}(t)=0$. $f^c_{ij}(t)$ can only equal one when $P^c_{ij}(t)\geq 1$. Thus, constraint (\ref{eqn:fcon}) is satisfied automatically for every slot under the proposed algorithm.

\begin{Theorem}\label{thm:capacity2}
For a network with given link rates $\{r_{ij}\}$ and a feasible activation pattern set $\Sc$, the network capacity region $\Lambda'$ for the constructed network consists of all rate matrices $\{\lambda^c_n\}$ for which there exists flow variables $\{\mu^{vc}_{ij}\}, v=1,2,3$ together with probabilities $\pi_s$ for all possible activation pattern $s\in S$ such that
\begin{align}
\mu^{vc}_{ij}&\geq 0,\quad \mu^{vc}_{ci}=0,\quad \mu^{vc}_{ii}=0, \quad \forall i,j,v,c\label{eqn:cap21}\\
\sum_{l}\mu^{3c}_{li}+\lambda^c_i&\leq \sum_{j}\mu^{1c}_{ij},\quad \forall i\neq c\label{eqn:cap22}\\
\mu^{1c}_{ij}&\leq \mu^{2c}_{ij},\quad \mu^{2c}_{ij}\leq \mu^{3c}_{ij}\quad \forall i,j,c\label{eqn:cap24}\\
\sum_c\mu^{2c}_{ij}&\leq \sum_c\sum_{s\in S}\pi_s\theta^c_{ij}(s)r_{ij},\quad \forall i,j\label{eqn:cap23}\\
\sum_{s\in \Sc}\pi_s&\leq1
\end{align}
where the probabilities $\theta^c_{ij}(s)$ satisfies
\begin{align}
\theta^c_{ij}(s)=0 \mbox{ if }i\notin s, \quad \sum_{c,j}\theta^c_{ij}(s)= 1,\forall i\label{eqn:cap25}
\end{align}
\end{Theorem}
\begin{Proof}
The necessary part can be proved in the same way for Theorem~\ref{thm1}. The sufficiency will be proved through constructing an algorithm that stabilizes all rate vectors satisfying the constraints.
\end{Proof}

In this constructed virtual network, $\mu_{ij}^{1c}, \mu_{ij}^{2c}, \mu_{ij}^{3c}$ can be interpreted as the flow over links $(i,v_{ij})$, $(v_{ij},p_{ij})$, $(p_{ij},j)$, respectively.
The constraints (\ref{eqn:cap21}) represent non-negativity and flow efficiency constraints. The constraints in (\ref{eqn:cap22}), (\ref{eqn:cap24}) represent flow conservation constraints, where the exogenous arrival flow rates for nodes $v_{ij}, p_{ij}$ are zero. The constraints in (\ref{eqn:cap23}) represent the physical link constraint for virtual link $(v_{ij},p_{ij})$, which equals the link constraint for the real link $(i,j)$ in the original system. Note that there is no explicit link constraints for $(i,v_{ij})$ and $(p_{ij},j)$, since the transfer of packets over these links happen at the same node, and there is no physical link constraint on them.

\begin{Lemma}
The network capacity region for the virtual network $\Lambda'$ defined in Theorem~\ref{thm:capacity2} is equal to that for the original system $\Lambda$ defined in Theorem~\ref{thm1}.
\end{Lemma}
\begin{Proof}
First, we show that if $\lambdav\in \Lambda' $, then it must lie in $\Lambda$ as well. This can be shown directly by letting $\mu^c_{ij}=\mu_{ij}^{2c}$, Thus, we have $\mu_{ij}^{1c}\leq \mu^c_{ij}\leq\mu_{ij}^{3c}$. Plugging into (\ref{eqn:cap22}), we have (\ref{eqn:cap2}), i.e., if $\lambdav$ satisfies the constraints in (\ref{eqn:cap21})-(\ref{eqn:cap25}), it must satisfy (\ref{eqn:cap1})-(\ref{eqn:cap5}) as well. Thus $\lambdav\in \Lambda$.

The other direction can be shown in the following way: we prove that for any $\lambdav+\epsilonv\in \Lambda$, $\lambdav+\frac{\epsilon}{2d+1}\in \Lambda'$, where $d$ is the maximum degree of the network.

Since $\lambdav+\epsilonv\in \Lambda$, we have
\begin{align}
\sum_{l}\mu^c_{li}+\lambda^c_i+\epsilon&\leq \sum_{j}\mu^c_{ij},\quad \forall i\neq c.\label{eqn:cap32}
\end{align}
By letting $\mu_{ij}^{2c}=\mu^c_{ij}$, we have that (\ref{eqn:cap23})-(\ref{eqn:cap25}) satisfied. At the same time, we let $\mu_{ij}^{1c}+\epsilon_1=\mu_{ij}^{3c}-\epsilon_1=\mu_{ij}^{2c}$, and plug them into (\ref{eqn:cap32}), which gives
\begin{align}
\sum_{l}(\mu^{3c}_{li}-\epsilon_1)+\lambda^c_i+\epsilon&\leq \sum_{j}(\mu^{1c}_{ij}+\epsilon_1),\quad \forall i\neq c.
\end{align}
Therefore,
\begin{align}
\sum_{l}\mu^{3c}_{li}+\lambda^c_i+\epsilon-2d\epsilon_1&\leq \sum_{j}\mu^{1c}_{ij},\quad \forall i\neq c.
\end{align}
By letting $\epsilon-2d\epsilon_1=\epsilon_1$, we have
\begin{align}
\sum_{l}\mu^{3c}_{li}+\lambda^c_i+\epsilon_1&\leq \sum_{j}\mu^{1c}_{ij},\quad \forall i\neq c\\
\mu_{ij}^{1c}+\epsilon_1&=\mu_{ij}^{2c},\quad \mu_{ij}^{2c}+\epsilon_1=\mu_{ij}^{3c}.
\end{align}
Thus, we have $\lambdav+\epsilonv_1\in \Lambda'$. As $\epsilon\rightarrow 0$, $\epsilon_1$ approaches zero as well. Thus, $\Lambda=\Lambda'$.
\end{Proof}


\begin{Theorem}\label{thm:virtual}
\tc{For $\eta\geq 1$}, the proposed algorithm stabilizes any rate vector satisfying $\lambdav+\epsilonv \in \Lambda$. The average expected queue backlog in the system is upper bounded by $$\frac{(2d+1)(KN(d+A_{max})^2+2N+(2\eta+1)KNd)}{\epsilon}.$$\end{Theorem}
The proof of the theorem is given in Appendix~\ref{apx:thm_virtual}. \tc{Since the upper bound is monotonically increasing in $\eta$, we can always set $\eta$ to 1 to achieve a better delay performance.}

The algorithm updates every slot. This avoids the delay caused by infrequent policy updating in the $T$-slot algorithm. On the other hand, we introduce virtual queues in the system. Since the differential backlog in Step 1) of the algorithm is not the physical differential backlog between nodes $(i,j)$ in the real system, the inaccuracy of the queue length information can, potentially, increase the average backlog in the system. This is reflected by the $2d+1$ factor in the upper bound. But for arrival rate vectors that are close to the boundary of network capacity region, the $T$-slot algorithm can only stabilize the system if $T$ is large. In such situation thus the virtual-queue based algorithm attains a better delay performance.

The algorithm exhausts the maximum stability region of the network without any pre-specified parameter depending on traffic statistics. Compared to the $T$-slot algorithm where the stabilizing parameter $T$ actually depends on how close a rate vector $\lambdav$ gets to the boundary of the network stability region $\Lambda$, this is a big advantage.

\section{Discussions}
\subsection{Enhanced Virtual Queue Based Algorithm}
According to Theorem~\ref{thm:Tslot}, the average expected backlog in the system under $T$-slot algorithm is $O(T^2)$, which indicates poor delay performance when $T$ is large. The virtual queue based algorithm avoids the long delay caused by the infrequent updating of queue length information. However, because of the introduction of virtual nodes, packets accumulate in virtual queues over weak links, which negatively impacts delay performance, especially when the system is lightly loaded. In order to improve delay performance, we proposed to enhance the virtual queue based algorithm by adjusting the term associated with $V_{ij}$ in the Lyapunov function.

Define a modified Lyaponov function
\begin{align}\label{eqn:lya3}
L(\Uv(t),\Vv(t),\Pv(t))&=\sum_{c,i}(U^c_i(t))^2+\sum_{c,(i,j)}(V^c_{ij}(t)+\gamma/ r_{ij})^2\nonumber\\
&\quad+\sum_{c,(i,j)}(P^c_{ij}(t)-\eta)^2.
\end{align}

Compared with (\ref{eqn:lya12}), we have added a term $\gamma/ r_{ij}$ to each $V^c_{ij}(t)$ in (\ref{eqn:lya3}). This is equivalent to add queue backlog $\gamma/ r_{ij}$ in the virtual queue. 
Following an analysis similar to that of the virtual queue algorithm (cf. Appendix~\ref{apx:lemma_1drift}), we can show that under constraints (\ref{eqn:bcon}) and (\ref{eqn:fcon}), the sample path Lyapunov drift satisfies
\begin{align*}
\Delta(t)&\leq \hspace{-0.02in}2\sum_{c,i}U^c_i(t)A^c_i(t)\hspace{-0.02in}-\hspace{-0.02in}2\hspace{-0.02in}\sum_{c,i,j}[V^c_{ij}(t)+\gamma/r_{ij}-P^c_{ij}(t)]\beta^c_{ij}(t)\nonumber\\
&\quad-2\sum_{c,i,j}[U^c_i(t)-(V^c_{ij}(t)+\gamma/r_{ij})]\beta^{1c}_{ij}(t)\nonumber\\
&\quad -2\sum_{c,i,j}[(P^c_{ij}(t)-\eta)-U^c_j(t)]f^c_{ij}(t)+\alpha_3
\end{align*}
where $\alpha_3$ is a positive constant. In order to minimize the one-step Lyapunov drift, we substitute the following ``modified'' step M1) for step 1) of the virtual queue based algorithm as follows:
\begin{itemize}
\item[M1)] \textbf{Find per-link backpressure.} At the beginning of a timeslot, node $i$ checks its neighbors and computes the differential backlogs. We compute the weight for the link between node $i$ and the first virtual node, and, separately, the weight for the link between the two virtual nodes. Specially, the weight for control action $\beta^{1c}_{ij}$ is computed as
\begin{align*}
W^{1c}_{ij}(t)= &[U^c_i(t)-(V^c_{ij}(t)+\gamma/r_{ij})\\
&+(V^c_{ij}(t)+\gamma/r_{ij}-P^c_{ij}(t))r_{ij}]^+,
\end{align*}
and the weight for control action $\beta^{2c}_{ij}$ is computed as
$$W^{2c}_{ij}(t)=[V^c_{ij}(t)+\gamma/r_{ij}-P^c_{ij}(t)]^+r_{ij}.$$
The weight of commodity $c$ over link $(i,j)$ is $W^c_{ij}(t)=\max\{W^{1c}_{ij}(t),W^{2c}_{ij}(t)\}$.
The weight for link $(i,j)$ is $$W_{ij}(t)=\max_c W^{c}_{ij}(t),$$ and the optimal commodity $$c^*_{ij}=\arg\max_c W^{c}_{ij}(t).$$
\end{itemize} 
The rest of the steps remain the same as in Sec.~\ref{sec:virtual_algo}. Following similar steps as in the proof of Theorem 4, we can show that the enhanced version also achieves the maximum stability region. The intuition is, in heavily backlogged regime $V^c_{ij}(t)\gg \gamma/r_{ij}$, therefore the added queue backlog is negligible and doesn't impact the stability of the queue.

As in the virtual queue based algorithm, in the enhanced algorithm, we also set $\eta=1$ to guarantee that $f^c_{ij}(t)=1$ only when $P^c_{ij}(t)\geq 1$. We now discuss the effect of the $\gamma$ parameter. We set $\gamma=1/2$. By setting $\gamma$ to be some positive value, the system adds some {\it virtual} backlog to buffer $V_{ij}$, thus preventing packets from entering the empty buffers over the weak links when the system starts from an initial empty state. It also increases backpressure between $V_{ij}$ and $P_{ij}$. Therefore, packets tend to be pushed through links more quickly, and the decoding time is shortened accordingly. Besides, in the modified Lyaponov function, we select weights for the {\it virtual} backlogs of virtual queues as the inverse of link rates. The reason for such selection is that the number of slots required to deliver a packet through a link is equal to the inverse of the link rate. We aim to capture different delay effects over different links through this adjustment. The intuition behind the enhanced algorithm is that, when the system is lightly loaded, passing packets only through strong links can support the traffic load while still providing good delay performance. Therefore, using weak links is not necessary, and using strong links is preferable. Setting the virtual backlog length to be $\gamma/r_{ij}$ forces packets to select strong links and improve the delay performance of the virtual queue based algorithm in the low traffic regime. When the system is heavily loaded and strong links cannot support all traffic flows, the differential backlogs over certain strong links eventually \tc{decrease}, and weak links start to be used. \tc{The enhanced algorithm essentially is a hybrid of the classic backpressure algorithm ($T=1$) and virtual queue based algorithm. It always stabilizes the system, and automatically adjusts the portion of slots that the system operates under each algorithm to maintain a good delay performance. }

\subsection{Dependent Link Rates}\label{sec:vary}
For the simplicity of analysis, in previous sections we assume that the link capacity for a given transmitter-receiver pair is fixed. We can easily relax this assumption and generalize the system model by assuming that the link rates are not fixed but are rather a function of the chosen activation pattern. Specifically, to be better able to capture the effects of interference, for link pair $(i,j)$ we assume the link rate under activation pattern $s$ is $r_{ij}(s)$. Then, the network capacity region under the new assumption is characterized by the same inequalities in Theorem~\ref{thm1} except that in eqn. (\ref{eqn:cap3}) $r_{ij}(s)$ is used instead of $r_{ij}$.

The scheduling algorithms should be adjusted accordingly in order to achieve the network capacity region. E.g., for the $T$-slot algorithm, in each updating slot, after determining the maximizing commodity $c^*_{ij}$ for each link $(i,j)$, the system selects the activation pattern, as well as the corresponding forwarder for each active transmitter, to maximize
\begin{align*}
  \max_{s}\sum_{i\in s,j\in\Nc(i)}[Q^{c^*}_i(t)-Q^{c^*}_j(t)]^+r_{ij}(s).
\end{align*}
The remaining steps remain the same.
For the virtual queue based algorithm, the maximizing commodity should be jointly selected with the activation pattern and the corresponding forwarders at the same time.

\subsection{Distributed Implementation}
The $T$-slot algorithm and the virtual queue based algorithm presented in the previous sections involving solving a constrained optimization problem (max-weight matching) in a centralized fashion. Here we consider distributed implementations. We assume nodes have information of the link rates between themselves and their neighbors, and the queue backlogs of their neighbors.

In interference-free networks where nodes can transmit simultaneously without interfering with each other, minimizing the Lyaponov drift function can be decomposed into local optimization problems. Each node individually makes the transmission decision based only on locally available information.

In networks with non-trivial interference constraints, we can use the message passing algorithm proposed in \cite{shah_msg_pass07} to select the activation pattern in a distributed way. First, each node selects its forwarder based on its local information. Then, the system uses the belief propagation based message-passing algorithm to select the optimal activation pattern that minimizes the Lyaponov drift. If the underlying interference graph is bipartite, \cite{shah_msg_pass07} shows that the the message-passing algorithm always converges to the optimal solution. \tc{Another possible approach is to develop carrier sensing based randomized scheduling algorithms. Carrier sensing based distributed algorithms are discussed in \cite{jiang, shah_random}, etc. The throughput optimality of these algorithms are established under certain assumptions. Some other distributed implementations of backpressure algorithms are discussed in Chapter 4.8 of \cite{Neely_now}. }

\section{Simulation Results}\label{sec:simu}
In this section we present detailed simulation results. These results
illustrate the basic properties of the algorithms.

\subsection{A single-commodity scenario}
\begin{figure}[t]
\begin{center}
\scalebox{0.45} {\epsffile{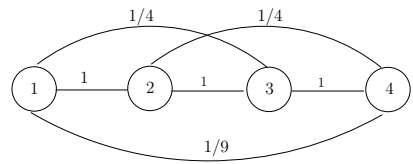}}
\end{center}
\vspace{-0.15in}
  \caption{The 4-node network used to compare the $T$-slot algorithm and the virtual queue based algorithm. The number labeling each link is the rate of that link.}
  \label{fig:4node}
\end{figure}

First consider the $4$-node wireless network shown in Fig.~\ref{fig:4node} where the links with nonzero rates are shown. We consider a single commodity scenario where new packets destined for node $4$ arrive at node $1$ and node $2$ according to independent Bernoulli processes with rate $\lambda_1$ and $\lambda_2$, respectively. Node $3$ does not have incoming packets. It acts purely as a relay. We assume that the system does not have any activation constraints, i.e., all nodes can transmit simultaneously without causing each other interference.

\begin{figure}[t]
\begin{center}
\scalebox{0.5} {\epsffile{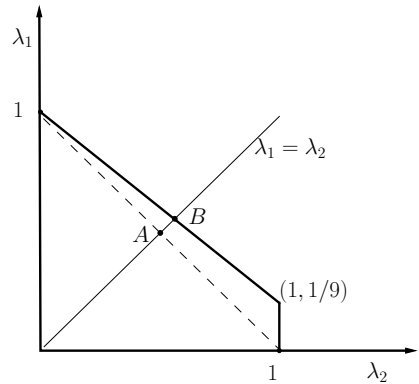}}
\end{center}
\vspace{-0.15in}
  \caption{The maximum stability region of the 4-node network. The inside triangular region is the network capacity region when MIA is not exploited.}
  \label{fig.stability_region}
\end{figure}

The maximum stability region $\Lambda$, shown in Fig.~\ref{fig.stability_region}, is the union of rate pairs ($\lambda_1$, $\lambda_2$) defined according to Theorem~\ref{thm1}. If mutual information accumulation is not allowed, the corresponding maximum stability region is indicated by the dashed triangle inside $\Lambda$. This follows because when weak links are not utilized, the only route from node 1 to node 4 is through node 2, thus the sum of arrival rates from node 1 and 2 cannot exceed link rate 1. When mutual information accumulation is exploited, the weak link from node 1 to node 4 with rate $1/9$ can be utilized, expanding the stability region.

\begin{figure}[t]
\begin{center}
\scalebox{0.3} {\epsffile{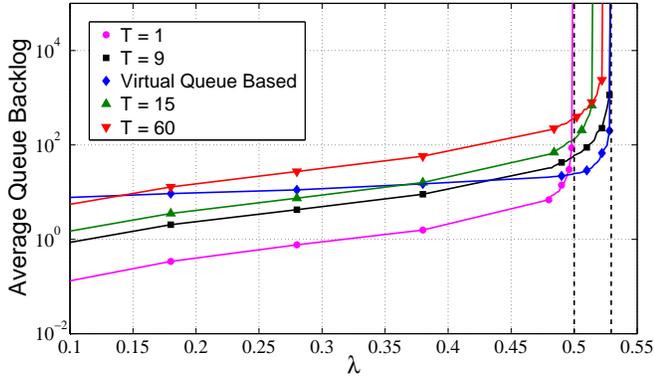}}
\end{center}
\vspace{-0.15in}
  \caption{Comparison of average backlog in the system under the algorithms.}
  \label{fig:avgbacklog}
\end{figure}

We first compare the performance of the $T$-slot algorithm for different values of $T$.  For each $T$, we conduct the simulation for arrival rates $\lambda_1=\lambda_2=\lambda$ ranging from 0 to 0.55. The resulting average backlog curve is shown in Fig.~\ref{fig:avgbacklog}. When $T=1$, the weak links cannot be utilized, and the algorithm can only stabilize the arrival rates up to $\lambda=1/2$, which is point $A$ in Fig.~\ref{fig.stability_region}. When $T=9$, the reciprocal of the link rate 1/9, the algorithm can stabilize arrival rates up to $\lambda=9/17$, corresponding to point $B$ in Fig.~\ref{fig.stability_region}. In this case, all of the partial packets transferred over weak link $(1,4)$ are eventually decoded, and that weak link is fully utilized. This is a special scenario since the value of $T$ perfectly matches the rate of weak link. For larger networks that consist of many weak links, selecting $T$ to match all of the weak links may be prohibitive, since such value can be very large. Except for such perfect matching scenarios, for more general values of $T$, the weak link is partially utilized, therefore, the maximum $\lambda$ the algorithm can stabilizes is some value between 1/2 and 9/17. In general, a larger $T$ stabilizes larger arrival rates, and results in an increased average backlog in the system. This is illustrated by curves with $T=15,60$ in Fig.~\ref{fig:avgbacklog}. 


Fig.~\ref{fig:avgbacklog} also plots the performance of the virtual queue based algorithm. The system can be stabilized up to $\lambda=9/17$ under the virtual queue based algorithm. Compared with the $T$-slot algorithm, the virtual queue based algorithm attains a much better delay performance for large value of $\lambda$, i.e., when the system is in heavy traffic regime. It dominates even the curve with $T=9$ at high rates. For small values of $\lambda$, the virtual queue based algorithm has a worse performance in terms of delay. This is because the algorithm requires the virtual queues to build up certain lengths in order to push packets through the weak links. The virtual queue based algorithm has relatively constant delay performance for $\lambda\in [0,1/2]$, while under the $T$-slot algorithm, the average backlog increases monotonically with $\lambda$.

\subsection{A multi-commodity scenario}
\begin{figure}[t]
\begin{center}
\scalebox{0.45} {\epsffile{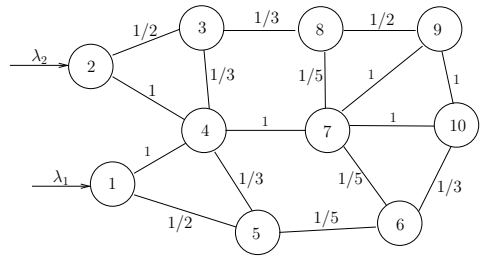}}
\end{center}
  \caption{The 10-node network used to compare the $T$-slot algorithm and the virtual queue based algorithm. The number labeling each link is the rate of that link.}
  \label{fig:10node}
\end{figure}

Next we consider the $10$-node network shown in Fig.~\ref{fig:10node}. We consider a multi-commodity scenario in which packets destined for node $10$ arrive at node $1$ and packets destined for node $9$ arrive at node $2$. Arrivals are distributed according to two independent Bernoulli processes with rate $\lambda_1$ and $\lambda_2$, respectively. We assume that the system does not have any activation constraints so that all nodes can transmit simultaneously without causing interference.

\begin{figure}[t]
\begin{center}
\scalebox{0.5} {\epsffile{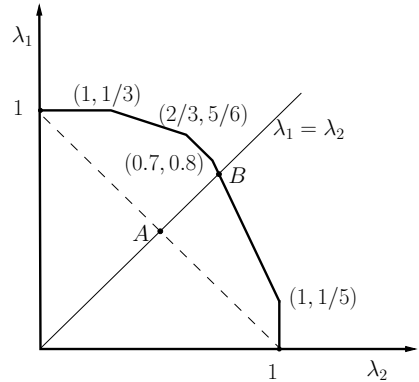}}
\end{center}
  \caption{The maximum stability region of the 10-node network, where $B=(11/15,11/15)$. The dashed triangular region is the network capacity region when MIA is not exploited.}
  \label{fig.stability_region2}
\end{figure}

The maximum stability region $\Lambda$ is shown in Fig.~\ref{fig.stability_region2}. If mutual information accumulation is not allowed, the corresponding maximum stability region is the dashed triangle inside $\Lambda$. This follows because when weak links are not used, the routes from node 1 to node 10 and the routes from node 2 to node 9 must pass through link $(4,7)$, thus the sum of arrival rates from node 1 and 2 cannot exceed that link rate. When mutual information accumulation is exploited, weak links can be utilized and form additional routes $1\rightarrow 5\rightarrow 6\rightarrow 10$, $2\rightarrow 3\rightarrow 8\rightarrow 9$, thus an expanded stability region can be achieved.

\begin{figure}[t]
\begin{center}
\scalebox{0.5} {\epsffile{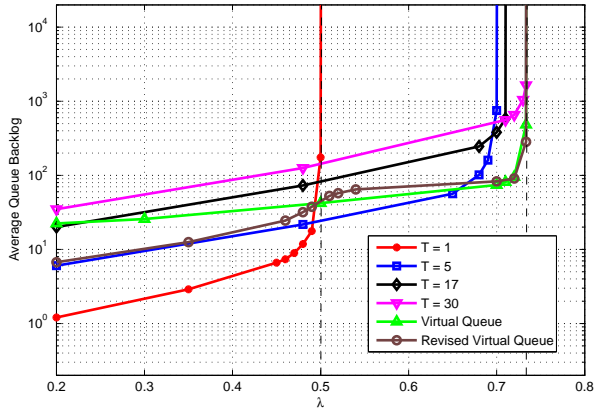}}
\end{center}
\vspace{-0.15in}
  \caption{Comparison of average backlog in the system under the algorithms.}
  \label{fig:avgbacklog2}
\vspace{-0.15in}
\end{figure}

We first compare the performance of the $T$-slot algorithm for different values of $T$.  For each $T$, we conduct the simulation for arrival rates $\lambda_1=\lambda_2=\lambda$ ranging from 0 to 0.75. The resulting average backlog curve is shown in Fig.~\ref{fig:avgbacklog2}. When $T=1$, the weak links are not utilized, so the algorithm can only stabilizes the arrival rates up to $\lambda=1/2$, which is point $A$ in Fig.~\ref{fig.stability_region2}. When $T=30$, which is the reciprocal of the link rate product $\frac{1}{2}\frac{1}{3}\frac{1}{5}$ and perfectly matches the rates of weak links, the algorithm can stabilize the arrival rates up to $\lambda=11/15$, corresponding to point $B$ in Fig.~\ref{fig.stability_region2}. For more general values of $T$, the weak links are partially utilized, therefore, the maximum $\lambda$ the algorithm can stabilizes is some value between 1/2 and 11/15. In general, a larger $T$ stabilizes larger arrival rates, and results in an increased average backlog in the system. This is illustrated by curves with $T=5,17$ in Fig.~\ref{fig:avgbacklog2}. In order to achieve the boundary point, a large $T$ is required in general.

In Fig.~\ref{fig.stability_region2}, we also present performance results from simulation of the virtual queue based algorithm. As expected, the system can be stabilized up to the edge of the stability region, $\lambda=11/15$. Compared with the $T$-slot algorithms, the virtual queue based algorithm attains a much better delay performance in heavy traffic regime. It dominates the curve with $T=30$ over the displayed rate region. Similar to the single-commodity scenario, for small values of $\lambda$, the virtual queue based algorithm does not show much advantage in terms of delay. Finally, we also provide simulation results for the enhanced virtual queue based algorithm in Fig.~\ref{fig:avgbacklog2}. The enhanced algorithm stabilize the full sepctrum of arrival rate, i.e., every input rate vector up to $\lambda=11/15$. The delay performance in the light traffic regime ($\lambda<1/2$) is improved under the enhanced version, with a smaller penalty of delay in the heavy traffic regime. The delay performance transition around $\lambda<1/2$ can be explained by the \tc{hybrid of and automatic adjustment between the classic backpressure ($T=1$) and virtual queue based algorithm. In the simulation, we set $\gamma=1/2$. The value of $\gamma$ decides the tradeoff between the delay performance in the light traffic regime and the delay performance in the heavy traffic regime.   }

\section{Conclusions}\label{sec:conclusions}
In this paper, we analyzed the optimal routing and scheduling policy when mutual information accumulation is exploited at the physical layer in a wireless network. We first characterized the maximum stability region under the natural assumption of policy space, which is shown to surpass the network capacity region when mutual information accumulation is not allowed. Two scheduling policies are proposed to cope with the decoding process introduced by mutual information accumulation. The $T$-slot algorithm and the virtual queue based algorithm can both achieve the maximum stability region but the latter has significantly reduced delay in heavy traffic regime. We also compared the performance under these two policies analytically and numerically.

\appendices
\section{Proof of the Necessity Part of Theorem~\ref{thm1}}\label{apx:thm1}
Suppose that a stabilizing control strategy exists. It is possible that the strategy may use redundant packet transfers, i.e., allowing multiple copies of a packet to exist at the same time in the network. However, under assumption A3, each packet can only have a single parent, i.e., once a node starts to accumulate information for a packet, it can only decode the packet if the total received information from a single transmitter exceeds the amount of information contained in that packet.

Define $X^c_i(t)$ as the total number of packets of commodity $c$ that have arrived at node $i$ up to slot $t$. Define $\Dc^c(t)$ as the set of distinct packets of commodity $c$ delivered to the destination node $c$ over $[0,t)$, and $D^c(t)=|\Dc^c(t)|$ be the total number of such distinct packets. Then, we have $D^c(t)=\sum_{i=1}^n Y^c_i(t)$, where $Y^c_i(t)$ is defined in Section~\ref{sec:capacity}. If multiple copies of a packet depart from the system, we only count the first delivered copy. In a stable system, for any traffic flow of commodity $c$ entering source node $i$, the average number of distinct packets delivered to destination node $c$ in each slot must be equal to its arrival rate, thus, we have
\begin{align}\label{eqn:consv}
  \lim_{t\rightarrow \infty} \frac{Y^c_i(t)}{t}= \lim_{t\rightarrow \infty} \frac{X^c_i(t)}{t}=\lambda^c_i.
\end{align}

For each distinct delivered packet $k$, there is a single routing path this packet took from its origin node to its destination node. Denote $D^c_{ij}(t)$ as the total number of distinct packets in $\Dc^c(t)$ transmitted from node $i$ to node $j$ over $[0,t)$. We have
\begin{align}\label{eqn:flow}
  \sum_{l=1,l\neq i}^n D^c_{li}(t)+Y^c_i(t)&=  \sum_{j=1,j\neq i}^n D^c_{ij}(t).
\end{align}

Define $T_s(t)$ to be the number of slots within which the system operates with activation pattern $s\in \Sc$ up to time $t$, and let $T^c_{ij}(s,t)$ be the number of slots that link $(i,j)$ is active and transmitting a packet \tc{in $\Dc^c(t)$} under activation pattern $s$ up to time $t$. Therefore, we have
\begin{align}
 D^c_{ij}(t)&= \sum_{s\in \Sc}T^c_{ij}(s,t)r_{ij},
\end{align}
 Thus,
\begin{align}
\frac{ D^c_{ij}(t)}{t}&= \sum_{s\in \Sc}\frac{T_s(t)}{t}\frac{T^c_{ij}(s,t)}{T_s(t)}r_{ij}.
\end{align}
Define $\mu^c_{ij}(t)=\frac{D^c_{ij}(t)}{t}$, $\pi_s(t)=\frac{T_s(t)}{t}$ and $\theta^c_{ij}(s,t)=\frac{T^c_{ij}(s,t)}{T_s(t)}$. We note that since we can deliver at most one packet to $i$ from $j$ in each time slot,
\begin{align}\label{eqn:con1}
0\leq\mu^c_{ij}(t)\leq 1, \quad \mu^c_{ii}(t)=0,\quad \mu^c_{c,i}(t)=0.
\end{align}
Because only one activation pattern is allowed per slot, we have
\begin{align}
\sum_{s\in \Sc}\pi_s(t)&=1.
\end{align}
On the other hand, since a node can only transmit a single packet in any slot, and we only count distinct copies of packets, then, if node $i$ is transmitting a packet of commodity $c$ at time $t$, at most one of the received copies at its neighbors can be counted as a distinct packet in $\Dc^c(t)$. Thus, we have
\begin{align}
\sum_{c,j}\theta^c_{ij}(s,t)&\leq 1\quad \textrm{if }i\in s.\label{eqn:con2}
\end{align}
We can always make inequality (\ref{eqn:con2}) tight by restricting to the policy space that a node can only transmit when it is necessary, i.e., if node $i$'s transmission at time $t$ does not contribute to the delivery of any distinct packet to its destination, it should keep silent and be removed from the activation pattern at time $t$. The remaining active nodes form another valid activation pattern, and gives $T_s(t)=\sum_{c,j}T^c_{ij}(s,t)$ for every $i\in s$.

These constraints define a closed and bounded region with finite dimension, thus there must exist an infinite subsequence $\tilde{t}$ over which the individual terms converges to points $\mu^c_{ij}$, $\pi_s$ and $\theta^c_{ji}(s)$ that also satisfy the inequalities (\ref{eqn:con1})-(\ref{eqn:con2}):
\begin{align}
 \lim_{\tilde{t}\rightarrow \infty} \mu^c_{ij}({\tilde{t}})=\mu^c_{ij},\\
   \lim_{\tilde{t}\rightarrow \infty} \pi_s(\tilde{t})=\pi_s,\\
   \lim_{\tilde{t}\rightarrow \infty} \theta^c_{ij}(s,\tilde{t})=\theta^c_{ij}(s).
\end{align}
Furthermore, using (\ref{eqn:consv}) in (\ref{eqn:flow}) and taking $\tilde{t}\rightarrow \infty$ yields
\begin{align}
\sum_{l}\mu^c_{li}+\lambda^c_i&= \sum_{j}\mu^c_{ij},\quad \forall i\neq c.
\end{align}
This proves the result.


\section{Proof of Theorem~\ref{thm:Tslot}}\label{apx:thm_Tslot}
Our algorithm always transmit packets from the head of $Q^c_i(t)$, and there is at most one full copy of each packet in the network. Therefore, without worry of confusion, in the following analysis we drop packet index $k$ and use commodity index $c$ instead as the superscript of control actions $\{\beta^{(k)}_{ij}\}$ and $\{f^{(k)}_{ij}\}$.

First, define the Lyapunov function $$L(\Qv(t))=\sum_{c,i}(Q^c_i(t))^2,$$ and the $T$-slot sample path Lyapunov drift as $$\Delta_T(t):=L(\Qv(t+T))-L(\Qv(t)).$$ Then, we have the following Lemma.

\begin{Lemma}\label{lemma:drift}
Assume the system changes its policy every $T$ slots starting at $t=0$. For all $t=0, T, 2T,\ldots$ and all possible values of $\Qv(t)$, under a given policy $\{\beta^c_{ij}(t)\}$, we have
\begin{align}\label{eqn:lya2}
\Delta_T(t)&\leq \sum_{c,i}-2Q^c_i(t)\Big(T\sum_j \beta^c_{ij}(t) r_{ij}-T\sum_{l}\beta^c_{li}(t) r_{li}\nonumber\\&\quad-\sum_{\tau=0}^{T-1}A^c_i(t+\tau)-1\Big)+\alpha_1,
\end{align}
where 
\begin{align}
\alpha_1&= KNT^2(\mu_{max}+A_{max})^2+NT^2.
\end{align}
\end{Lemma}
\begin{Proof}
Under any given policy, the queue length evolves according to
\begin{align}
Q^c_i(t+1)&= \Big(Q^c_i(t)-\sum_{j}\beta^c_{ij}(t)f^c_{ij}(t)\Big)^+\nonumber\\
&\quad+\sum_{l}\beta^c_{li}(t)f^c_{li}(t)+A^c_i(t).\end{align}
Considering the policy which updates at $t=0,T,2T,\ldots$, we have
\begin{align}
&Q^c_i(t+T)\nonumber\\
&\leq \Big(Q^c_i(t)-\sum_{\tau=0}^{T-1}\sum_{j}\beta^c_{ij}(t)f^c_{ij}(t+\tau)\Big)^+\nonumber\\
&\quad+\sum_{\tau=0}^{T-1}\sum_{l}\beta^c_{li}(t)f^c_{li}(t+\tau)+\sum_{\tau=0}^{T-1}A^c_i(t+\tau)\label{eqn:sumT}\\
&\leq \Big(Q^c_i(t)-\sum_j \beta^c_{ij}(t)\lfloor r_{ij}T\rfloor\Big)^+\nonumber\\
&\quad+\sum_{l}\beta^c_{li}(t)\lfloor r_{li}T\rfloor+\sum_{\tau=0}^{T-1}A^c_i(t+\tau).\label{eqn:lya10}
\end{align}
In (\ref{eqn:sumT}), we upper bound $Q^c_i(t+T)$ by moving the negative terms into the function $(\cdot)^+$. This follows from the facts that
\begin{align*}
\max[a+b-c,0]&\leq \max[a-c,0]+b\\
\max[\max[a,0]-c,0]&=\max[a-c,0] \textrm{ for }a,b,c\geq 0.
\end{align*}
This is equivalent to letting node $i$ transmit packets existing in $Q^c_i(t)$ only, i.e., even if there are some packets of commodity $c$ arrive at node $i$ in the epoch and all of the packets existing in $Q^c_i(t)$ have been cleared, they are not transmitted until next epoch. Since under the policy, these packets may be transmitted to next hop, the actual queue length $Q^c_i(t+T)$ can be upper bounded. Eqn.~(\ref{eqn:lya10}) follows from the fact that over the T-slot window, the successfully delivered packets (including dummy packets) from node $i$ to node $j$ is $\lfloor T\beta^c_{ij}(t)r_{ij}\rfloor$, and recall $\beta^c_{ij}(t)$ is held constant for the whole epoch. Since both sides of the inequality are positive, it holds for the square of both sides, thus,
\begin{align}
&(Q^c_i(t+T))^2\nonumber\\
&\leq \Big(Q^c_i(t)-\sum_j \beta^c_{ij}(t)\lfloor r_{ij}T\rfloor\Big)^2\nonumber\\
&\quad+2Q^c_i(t)\Big(\sum_{l}\beta_{li}(t)\lfloor r_{li}T\rfloor+\sum_{\tau=0}^{T-1}A^c_i(t+\tau)\Big)\nonumber\\
&\quad+\Big(\sum_{l}\beta^c_{li}(t)\lfloor r_{li}T\rfloor+\sum_{\tau=0}^{T-1}A^c_i(t+\tau)\Big)^2\label{eqn:square}\\
&\leq(Q^c_i(t))^2-2Q^c_i(t)\Big(\sum_j \beta^c_{ij}(t)\lfloor r_{ij}T\rfloor\nonumber\\&\quad-\sum_{l}\beta^c_{li}(t)\lfloor r_{li}T\rfloor-\sum_{\tau=0}^{T-1}A^c_i(t+\tau)\Big)+ C^c_{i},
\end{align}
where
\begin{align*}
C^c_i&=\Big(\sum_{l}\beta^c_{li}(t)\lfloor r_{li}T\rfloor+\sum_{\tau=0}^{T-1}A^c_i(t+\tau)\Big)^2\\
&\quad+\Big(\sum_j \beta^c_{ij}(t)\lfloor r_{ij}T\rfloor\Big)^2.
\end{align*}
We use $Q^c_i(t)$ instead of $\left(Q^c_i(t)-\sum_j \beta^c_{ij}(t)\lfloor r_{ij}T\rfloor\right)^+$ for the cross term in (\ref{eqn:square}). Since the former is always greater than the latter, the inequality (\ref{eqn:square}) holds. Therefore, we have
\begin{align}
\Delta_T(t)&\leq \sum_{c,i} -2Q^c_i(t)\Big(\sum_j \beta^c_{ij}(t)\lfloor r_{ij}T\rfloor-\sum_{l}\beta^c_{li}(t)\lfloor r_{li}T\rfloor\nonumber\\&\quad-\sum_{\tau=0}^{T-1}A^c_i(t+\tau)\Big)+\sum_{c,i} C^c_{i}\\
&\leq  \sum_{c,i}-2Q^c_i(t)\Big(\sum_j \beta^c_{ij}(t)( r_{ij}T-1)-\sum_{l}\beta^c_{li}(t) r_{li}T\nonumber\\&\quad-\sum_{\tau=0}^{T-1}A^c_i(t+\tau)\Big)+\sum_{c,i} C^c_{i}\label{eqn:floor}\\
&\leq \sum_{c,i}-2Q^c_i(t)\Big(T\sum_j \beta^c_{ij}(t) r_{ij}-T\sum_{l}\beta^c_{li}(t) r_{li}\nonumber\\&\quad-\sum_{\tau=0}^{T-1}A^c_i(t+\tau)-1\Big)+\sum_{c,i} C^c_{i}\label{eqn:floor2}
\end{align}
where $\Delta_T(t)$ is the $T$-slot Lyapunov drift, (\ref{eqn:floor}) follows from the fact that $x-1<\lfloor x\rfloor\leq x$, and (\ref{eqn:floor2}) follows from the fact that $\sum_j \beta^c_{ij}(t)\leq 1$. Based on the assumptions that $A_i^c(t)\leq A_{max}$, the maximum number of decoded packets at a node is upper bounded by $\mu_{max}$, and the constraint (\ref{beta_con}), we have 
\begin{align*}
\sum_{c,i} C^c_i&\leq KNT^2(\mu_{max}+A_{max})^2+NT^2:=\alpha_1
\end{align*}
The proof is completed.
\end{Proof}
%
\begin{Lemma}\label{lemma:min}
For a given $\Qv(t)$ on slot $t=0,T,2T,\ldots$, under the $T$-slot algorithm, the $T$-slot Lyapunov drift satisfies
\begin{align}
\Delta_T(t)&\leq \sum_{c,i}-2Q^c_i(t)\Big(T\sum_{j}\hat{\beta}^c_{ij}(t)r_{ij}-T\sum_{l}\hat{\beta}^c_{li}(t)r_{li}\nonumber\\
&\quad-\sum_{\tau=0}^{T-1}A^c_i(t+\tau)-1\Big)+\alpha_1,\label{eqn:lya22}
\end{align}
where $\{\hat{\beta}^c_{ij}(t)\}$ are any alternative (possibly randomized) decisions that satisfy (\ref{beta_con})-(\ref{beta_con2}).
Further more, we have
\begin{align}
&\Ed\{\Delta_T(t)|\Qv(t)\}
\leq \Ed\Big\{\sum_{c,i}-2Q^c_i(t)\Big(T\sum_{j}\hat{\beta}^c_{ij}(t)r_{ij}\nonumber\\&\qquad\qquad-T\sum_{l}\hat{\beta}^c_{li}(t)r_{li}
\left.-T\lambda^c_i-1\Big)\right|\Qv(t)\Big\}+\alpha_1\label{eqn:lya4}
\end{align}
\end{Lemma}
\begin{Proof}
 Given $\Qv(t)$, the $T$-slot algorithm makes decisions to minimize the right hand side of (\ref{eqn:lya2}). Therefore, inequality (\ref{eqn:lya22}) holds for all realizations of the random quantities, and hence also holds when taking conditional expectations of both sides. Thus, we have (\ref{eqn:lya4}). 
\end{Proof}

\begin{Corollary}\label{cor1}
A rate matrix $\{\lambdav+\epsilonv\}$ is in the capacity region $\Lambda$ if and only if there exists a stationary (possibly randomized) algorithm that chooses control decisions subject to constraint (\ref{beta_con})-(\ref{beta_con2}) and independent of current queue backlog to yield
\begin{align}\label{eqn:stable}
\Ed\Big\{\sum_{j}\beta^c_{ij}r_{ij}-\sum_{l}\beta^c_{li}r_{li}-\lambda^c_i\Big\}\geq \epsilon\quad \forall i\neq c.
\end{align}
\end{Corollary}
\begin{Proof}
The result is an immediate consequence of Theorem~\ref{thm1}. The intuition is to think $\Ed\left\{\beta^c_{ij}\right\}r_{ij}$ as $\mu^c_{ij}$ in (\ref{eqn:cap1})-(\ref{eqn:cap5}). The necessary part is obtained directly. The sufficient part will be shown in the following section.
\end{Proof}

For any $\{\lambdav+\epsilonv\}\in \Lambda$, Lemma~\ref{lemma:min} shows that the $T$-slot algorithm minimizes the right hand side of (\ref{eqn:lya4}) for any alternative policy satisfying (\ref{beta_con})-(\ref{beta_con2}). On the other hand, Corollary~\ref{cor1} implies that such policy can be constructed in a randomized way which is independent of current queue status in the network and satisfying (\ref{eqn:stable}).
Combining Lemma~\ref{lemma:min} and Corollary~\ref{cor1}, we have
\begin{align*}
&\Ed\{\Delta_T(t)|\Qv(t)\}\\
&\leq \sum_{c,i}-2Q^c_i(t)\Big(\Big(\hat{\beta}^c_{ij}(t)r_{ij}-\sum_{l}\hat{\beta}^c_{li}(t)r_{li}-\lambda^c_i\Big)T-1\Big)+\alpha_1\\
&\leq \sum_{c,i}-2Q^c_i(t)\left(\epsilon T-1\right)+\alpha_1
\end{align*}
Taking expectations of the above inequality over the distribution of $\Qv(t)$, we have
\begin{align}
\Ed\{\Delta_T(t)\}&\leq \sum_{c,i}-2\Ed\{Q^c_i(t)\}\left(\epsilon T-1\right)+\alpha_1.
\end{align}
Summing terms over $t=0,T,\ldots,(M-1)T$ for positive integer $M$ yields
\begin{align*}
&\frac{\Ed\{L(\Qv(MT))-L(\Qv(0))\}}{M}\\
&\leq \frac{\sum_{m=0}^{M-1}\sum_{c,i}-2\Ed\{Q^c_i(mT)\}\left(\epsilon T-1\right)}{M}+\alpha_1,
\end{align*}
i.e.,
\begin{align}
\sum_{m=0}^{M-1}\sum_{c,i}\Ed\{Q^c_i(mT)\}&\leq \frac{ L(\Qv(0))-L(\Qv(MT))+M\alpha_1}{2\left(\epsilon T-1\right)}\nonumber\\
&\leq \frac{ L(\Qv(0))+M\alpha_1}{2\left(\epsilon T-1\right)}.\label{eqn:mT}
\end{align}
We drop $L(\Qv(MT))$ in (\ref{eqn:mT}) since $L(\Qv(MT))\geq 0$ based on the definition of Lyaponov function.
On the other hand, for $t=0,T,2T,\ldots$ and $0<\tau<T$, we have
\begin{align}
Q^c_i(t+\tau)\leq Q^c_i(t)+(\mu_{max}+A_{max})\tau
\end{align}
where $A_{max}$ is the maximum arrival rates and $\mu_{max}$ is the maximum number of decoded packets in a slot for any node. Therefore,
\begin{align*}
\sum_{m=0}^{M-1}\sum_{\tau=1}^{T-1}\sum_{c,i}&\Ed\{Q^c_i(mT+\tau)\}\leq \sum_{m=0}^{M-1}\sum_{c,i}(T-1)\Ed\{Q^c_i(mT)\}\nonumber\\
&\quad+MT(T-1)KN(\mu_{max}+A_{max})/2
\end{align*}
Combining with (\ref{eqn:mT}), we have
\begin{align*}
&\frac{1}{MT}\sum_{m=0}^{M-1}\sum_{\tau=0}^{T-1}\sum_{c,i}\Ed\{Q^c_i(mT+\tau)\}\nonumber\\
&\leq \frac{1}{M}\sum_{m=0}^{M-1}\sum_{c,i}\Ed\{Q^c_i(mT)\}+KN(T-1)(\mu_{max}+A_{max})/2\\
&\leq \frac{ L(\Qv(0))+M\alpha_1}{2M\left(\epsilon T-1\right)}+\frac{KN(T-1)(\mu_{max}+A_{max})}{2}
\end{align*}
Letting $M\rightarrow \infty$, we have
\begin{align*}
&\lim_{t\rightarrow \infty}\frac{1}{t}\sum_{\tau=0}^{t-1}\sum_{c,i}\Ed\{Q^c_i(\tau)\}\nonumber\\
&\leq \frac{\alpha_1}{2(\epsilon T-1)}+\frac{KN(T-1)(\mu_{max}+A_{max})}{2}
\end{align*}

Therefore, when $T> \frac{1}{\epsilon}$, the average expected backlog in the network is bounded. Thus, the system is stable.
Since our algorithm operates under assumptions A1-A2, the total number of partial packets in the system, $\sum_{c,i}P^c_i(t)$, is always upper bounded by $\sum_{c,i}{Q^c_i(t)}$. Therefore, if the $\{Q^c_i(t)\}$ are stable, the $\{P^c_i(t)\}$ must be stable, and the overall system is stable.

\section{Proof of Lemma~\ref{lemma:1drift}}\label{apx:lemma_1drift}
Based on (\ref{eqn:u})-(\ref{eqn:p}), we have
\begin{align}
(U^c_i(t+1))^2&\leq (U_i^c(t))^2+\Big(\sum_{j}\beta_{ij}^{1c}(t)\Big)^2\nonumber\\
&-2U^c_i(t)\Big(\sum_{j}\beta_{ij}^{1c}(t)-\sum_{l}f^c_{li}(t)-A^c_i(t)\Big)\nonumber\\
&\quad +\Big(\sum_{l}f^c_{li}(t)+A^c_i(t)\Big)^2\label{eqn:u2}\\
(V_{ij}^c(t+1))^2&\leq (V^c_{ij}(t))^2+2V^c_{ij}(t)\left(\beta_{ij}^{1c}(t)-\beta_{ij}^{c}(t)r_{ij}\right)\nonumber\\
&\quad+\left(\beta_{ij}^{1c}(t)-\beta_{ij}^{c}(t)r_{ij}\right)^2\label{eqn:v2}\\
(P_{ij}^c(t+1))^2&\leq (P^c_{ij}(t))^2+2P^c_{ij}(t)(\beta^c_{ij}(t)r_{ij}(t)-f^c_{ij}(t))\nonumber\\
&\quad+(\beta^c_{ij}(t)r_{ij}(t)-f^c_{ij}(t))^2\label{eqn:p2}\\
P^c_{ij}(t+1)&\geq P^c_{ij}(t)-f^c_{ij}(t)
\end{align}
Thus,
\begin{align*}
\Delta(t)&\leq \sum_{c,i}-2U^c_i(t)\Big( \sum_{j}\beta_{ij}^{1c}(t)-\sum_{l}f^c_{li}(t)-A^c_i(t)\Big)\nonumber\\
&\quad+\sum_{c,i,j}2V^c_{ij}(t)\left(\beta_{ij}^{1c}(t)-\beta_{ij}^{c}(t)r_{ij}\right)\nonumber\\
&\quad+\sum_{c,i,j}[2P^c_{ij}(t)(\beta^c_{ij}(t)r_{ij}(t)-f^c_{ij}(t))+2\eta f^c_{ij}(t)]+C
\end{align*}
where
\begin{align*}
C&=\sum_{c,i}\Big( \sum_{j}\beta_{ij}^{1c}(t)\Big)^2+\sum_{c,i}\Big(\sum_{l}f^c_{li}(t)+A^c_i(t)\Big)^2\\
&\quad+\sum_{c,i,j}\Big[\left(\beta_{ij}^{1c}(t)-\beta_{ij}^{c}(t)r_{ij}\right)^2+(\beta^c_{ij}(t)r_{ij}(t)-f^c_{ij}(t))^2\Big]
\end{align*}
Because of constraints (\ref{eqn:bcon})-(\ref{eqn:fcon}), we have
\begin{align}
C&\leq KN(d+A_{max})^2+2N+KNd:=\alpha_2.
\end{align}
Combining items with respect to link $(i,j)$, we have (\ref{eqn:delta}).

\section{Proof of Theorem~\ref{thm:virtual}}\label{apx:thm_virtual}
\begin{Corollary}\label{cor:f}
A rate vector $\lambdav+\epsilonv$ is in the capacity region $\Lambda'$ if and only if there exists a stationary (possibly randomized) algorithm that chooses control decisions (independent of current queue backlog) subject to constraints (\ref{eqn:bcon}), to yield
\begin{align}
\Ed\{\beta^c_{ij}\}r_{ij}&=\Ed\{\beta^{1c}_{ij}\}+\epsilon\\
\Ed\{f^c_{ij}\}&=\Ed\{\beta^c_{ij}\}r_{ij}+\epsilon\\
\Ed\Big\{\sum_{j}f^c_{ij}-\sum_{l}\beta^{1c}_{li}-\lambda^c_i\Big\}&\geq \epsilon\quad \forall i\neq c
\end{align}
\end{Corollary}

\begin{Proof}
The result is an immediate consequence of Theorem~\ref{thm:capacity2}. The intuition is to let $\Ed\{\beta^c_{ij}\}r_{ij}=\mu_{ij}^{2c}$, $\Ed\{\beta^{1c}_{ij}\}=\mu_{ij}^{1c}$ and $\Ed\{f^c_{ij}\}=\mu_{ij}^{3c}$. The necessary part is obtained directly. The sufficient part will be shown in the following proof.
\end{Proof}

\begin{Lemma}\label{lemma:virtual_drift}
Under the virtual queue based algorithm, 
\begin{align}
&\sum_{c,i,j}\Big\{[U^c_i(t)-V^c_{ij}(t)]\beta^{1c}_{ij}(t)+[V^c_{ij}(t)-P^c_{ij}(t)]r_{ij}\beta^c_{ij}
(t)\nonumber\\
&\quad +[P^c_{ij}(t)-\eta-U^c_j(t)]f^c_{ij}(t)\Big\}\nonumber\\
&\geq \sum_{c,i,j}\Big\{[U^c_i(t)-V^c_{ij}(t)]\hat{\beta}^{1c}_{ij}(t)+[V^c_{ij}(t)-P^c_{ij}(t)]r_{ij}\hat{\beta}^c_{ij}(t)\nonumber\\
&\quad +[P^c_{ij}(t)-\eta-U^c_j(t)]\hat{f}^c_{ij}(t)\Big\}\label{eqn:minVirtual}
\end{align}
for any other binary control policy $\{\hat{\beta}^{1c}_{ij},\hat{\beta}^c_{ij}, \hat{f}^c_{ij}\}$ satisfying (\ref{eqn:bcon}).
\end{Lemma}
\begin{Proof}
This lemma is an immediate consequence of the fact that the virtual queue based algorithm maximizes the left hand side of (\ref{eqn:minVirtual}) while satisfying (\ref{eqn:bcon}). The constraint (\ref{eqn:fcon}) is satisfied automatically.
\end{Proof}

Based on Lemma~\ref{lemma:1drift} (\ref{eqn:delta}) and Lemma~\ref{lemma:virtual_drift}, we have
\begin{align*}
&\Ed\{\Delta(t)|\Uv(t),\Vv(t),\Pv(t)\}\\&\leq -2\sum_{c,i,j}\Ed\left\{[U^c_i(t)-V^c_{ij}(t)]\hat{\beta}^{1c}_{ij}+[V^c_{ij}(t)-P^c_{ij}(t)]r_{ij}\hat{\beta}^c_{ij}\right.\\
&\left.\left.\quad +[P^c_{ij}(t)-\eta-U^c_j(t)]\hat{f}^c_{ij}(t)\right|\Uv(t),\Tv(t),\Pv(t)\right\}\\
&\quad+2\sum_{c,i}U^c_i(t)\lambda^c_i+\alpha_2\\
&\leq \Ed\Big\{\sum_{c,i}-2U^c_i(t)\Big( \sum_{j}\hat{\beta}_{ij}^{1c}(t)-\sum_{l}\hat{f}^c_{li}(t)-\lambda^c_i(t)\Big)\Big\}\\
&\quad+\Ed\Big\{\sum_{c,i,j}2V^c_{ij}(t)\left(\hat{\beta}_{ij}^{1c}(t)-\hat{\beta}_{ij}^{c}(t)r_{ij}\right)\Big\}\nonumber\\
&\quad+\Ed\Big\{\sum_{c,i,j}2P^c_{ij}(t)(\hat{\beta}^c_{ij}(t)r_{ij}(t)-\hat{f}^c_{ij}(t))+2\eta \hat{f}^c_{ij}(t)\Big\}+\alpha_2
\end{align*}

Since for $\lambdav+\epsilonv \in \Lambda$, we have $\lambdav+\epsilonv/(2d+1) \in \Lambda'$, thus 
\begin{align*}
&\Ed\{\Delta(t)|\Uv(t),\Vv(t),\Pv(t)\}\\
&\leq -2\Big(\sum_{c,i}U^c_i(t)+\sum_{c,i,j}V^c_{ij}(t)+\sum_{c,i,j}P^c_{ij}(t)\Big)\frac{\epsilon}{2d+1}\\
&\quad+2\eta KNd+KN(d+A_{max})^2+2N+KNd.
\end{align*}
Therefore, the system is stable, and the average backlog is upper bounded by
\begin{align*}
&\lim_{t\rightarrow\infty}\frac{1}{t}\sum_{\tau=0}^{t-1}\Big(\sum_{c,i}U^c_i(\tau)+\sum_{c,i,j}V^c_{ij}(\tau)+\sum_{c,i,j}P^c_{ij}(\tau)\Big)\\
&\leq \frac{(2d+1)(KN(d+A_{max})^2+2N+(2\eta+1)KNd)}{\epsilon}
\end{align*}
The proof is completed.
\bibliographystyle{IEEEtran}
\bibliography{scheduling}

\begin{thebibliography}{10}
\providecommand{\url}[1]{#1}
\csname url@samestyle\endcsname
\providecommand{\newblock}{\relax}
\providecommand{\bibinfo}[2]{#2}
\providecommand{\BIBentrySTDinterwordspacing}{\spaceskip=0pt\relax}
\providecommand{\BIBentryALTinterwordstretchfactor}{4}
\providecommand{\BIBentryALTinterwordspacing}{\spaceskip=\fontdimen2\font plus
\BIBentryALTinterwordstretchfactor\fontdimen3\font minus
  \fontdimen4\font\relax}
\providecommand{\BIBforeignlanguage}[2]{{%
\expandafter\ifx\csname l@#1\endcsname\relax
\typeout{** WARNING: IEEEtran.bst: No hyphenation pattern has been}%
\typeout{** loaded for the language `#1'. Using the pattern for}%
\typeout{** the default language instead.}%
\else
\language=\csname l@#1\endcsname
\fi
#2}}
\providecommand{\BIBdecl}{\relax}
\BIBdecl

\bibitem{ExOR}
S.~Biswas and R.~Morris, ``{ExOR}: {Opportunistic} multi-hop routing for
  wireless networks,'' \emph{ACM SIGCOMM}, vol.~35, pp. 133--144, Aug. 2005.

\bibitem{Rozner_2009}
E.~Rozner, J.~Seshadri, Y.~A. Mehta, and L.~Qiu, ``{SOAR}: {Simple}
  opportunistic adaptive routing protocol for wireless mesh networks,''
  \emph{IEEE Tran.\ Mobi.\ Comp.}, vol.~8, pp. 1622--1635, Dec. 2009.

\bibitem{Neely_2008}
M.~J. Neely and R.~Urgaonkar, ``Optimal backpressure routing for wireless
  networks with multi-receiver diversity,'' \emph{Proc.\ Conf.\ Info.\ Sci.\
  Sys.}, pp. 18--25, Jan. 2006.

\bibitem{Tassiulas_1992}
L.~Tassiulas and A.~Ephremides, ``Stability properties of constrained queueing
  systems and scheduling policies for maximum throughput in multihop radio
  networks,'' \emph{IEEE Trans.\ Automat.\ Contr.}, vol.~37, pp. 1936--1948,
  Apr. 1992.

\bibitem{Tassiulas_1993}
------, ``Dynamic server allocation to parallel queues with randomly varying
  connectivity,'' \emph{IEEE Trans.\ Inform.\ Theory}, vol.~39, pp. 466--478,
  Jun. 1993.

\bibitem{Yeh_2007}
E.~M. Yeh and R.~A. Berry, ``Throughput optimal control of cooperative relay
  networks,'' \emph{IEEE Trans.\ Inform.\ Theory}, vol.~53, pp. 3827--3833,
  Oct. 2007.

\bibitem{Draper_2008}
S.~C. Draper, L.~Liu, A.~F. Molisch, and J.~S. Yedidia, ``Cooperative routing
  for wireless networks using mutual-information accumulation,'' \emph{IEEE
  Trans.\ Inform.\ Theory}, vol.~57, no.~8, pp. 5757--5762, 2011.

\bibitem{Urgaonkar_2010}
R.~Urgaonkar and M.~J. Neely, ``Routing with mutual information accumulation in
  wireless networks,'' Tech. Rep., Aug. 2010.

\bibitem{mia_allerton11}
Y.~Liu, J.~Yang, and S.~C. Draper, ``Exploiting route diversity in multi-packet
  transmission using mutual information accumulation,'' \emph{Allerton Conf. on
  Commun., Control, and Computing}, Sept. 2011.

\bibitem{Maric_2004}
I.~Maric and R.~D. Yates, ``Cooperative multihop broadcast for wireless
  networks,'' \emph{IEEE J.\ Select.\ Areas Commun.}, vol.~22, pp. 1080--1088,
  Aug. 2004.

\bibitem{Maric_2005}
------, ``Cooperative multicast for maximum network lifetime,'' \emph{IEEE J.\
  Select.\ Areas Commun.}, vol.~23, pp. 127--135, Jan. 2005.

\bibitem{Chen_2005}
J.~Chen, L.~Jia, X.~Liu, G.~Noubir, and R.~Sundaram, ``Minimum energy
  accumulative routing in wireless networks,'' \emph{Proc.\ IEEE INFOCOM},
  vol.~3, pp. 1875--1886, Mar. 2005.

\bibitem{Mitzenmacher_2004}
M.~Mitzenmacher, ``Digital foutains: {A} survey and look forward,''
  \emph{Proc.\ IEEE Inform.\ Theory Workshop}, pp. 271--276, Oct. 2004.

\bibitem{LT_Code}
M.~Luby, ``{LT} codes,'' \emph{Proc.\ Symp.\ Foundation of Computer Science},
  pp. 271--282, 2002.

\bibitem{cohen}
E.~Yeh and A.~Cohen, ``Throughput optimal power and rate control for queued
  multiaccess and broadcast communications,'' \emph{IEEE International
  Symposium on Information Theory}, p. 112, June 2004.

\bibitem{Neely05}
M.~Neely, E.~Modiano, and C.Rohrs, ``Dynamic power allocation and routing for
  time-varying wireless networks,'' \emph{Selected Areas in Communications,
  IEEE Journal on}, vol.~23, pp. 89--103, Jan 2005.

\bibitem{Neely_now}
L.~Georgiadis, M.~Neely, and L.~Tassiulas, \emph{Resource Allocation and
  Cross-Layer Control in Wireless Networks}.\hskip 1em plus 0.5em minus
  0.4em\relax Foundations and Trends in Networking, 2006, vol.~1.

\bibitem{stable94}
W.~Szpankowski, ``Stability conditions for some multiqueue distributed systems:
  Buffered random access systems,'' \emph{Advances in Applied Probability},
  vol.~26, p. 498–515, 1994.

\bibitem{shah_msg_pass07}
S.~Sanghavi, D.~Shah, and A.~Willsky, ``Message-passing for maximum weight
  independent set,'' \emph{Proceedings of NIPS}, 2007.

\bibitem{jiang}
L.~Jiang and J.~Walrand, ``A distributed csma algorithm for throughput and
  utility maximization in wireless networks,'' \emph{IEEE/ACM Trans. Netw.},
  vol.~18, no.~3, pp. 960--972, Jun. 2010.

\bibitem{shah_random}
D.~Shah and J.~Shin, ``Randomized scheduling algorithms for queueing
  networks,'' \emph{accepted to appear in Annals of Applied Probability}, 2011.

\end{thebibliography}
\end{document}